\documentclass[10pt,twocolumn,showpacs,preprintnumbers,amsmath,amssymb,aps,prl,longbibliography,superscriptaddress,nofootinbib]{revtex4-2}

\usepackage{array,braket,mathtools,siunitx}
\usepackage{bm}

\usepackage[unicode=true,
bookmarks=true,bookmarksnumbered=true,bookmarksopen=false,
breaklinks=true,pdfborder={0 0 0},backref=false,colorlinks=true]{hyperref}
\hypersetup{citecolor={blue},urlcolor={magenta}}
\usepackage[normalem]{ulem}

\usepackage{cleveref}
\crefname{appendix}{App.}{Apps.}
\crefname{equation}{Eq.}{Eqs.}
\crefname{figure}{Fig.}{Figs.}
\crefname{table}{Tab.}{Tabs.}
\crefname{section}{Sec.}{Secs.}

\usepackage{amsfonts}
\usepackage{amsmath}
\usepackage{multirow}
\usepackage{amssymb}
\usepackage{amsbsy}
\usepackage{graphicx}
\usepackage{epstopdf}
\usepackage{color}
\usepackage{braket} 
\usepackage{mathdots} 
\usepackage{booktabs}
\usepackage{indentfirst}
\usepackage{diagbox}
\usepackage{lipsum}
\usepackage{mwe}
\usepackage{hyperref}

\hypersetup{colorlinks=true, linkcolor=blue, anchorcolor=blue, citecolor=blue,urlcolor=blue}

\begin{document}
\title{Exponential U(1) Symmetry-Breaking Phase as a Disorder-Free Quantum Glass}

\author{Yu-Min Hu}
\affiliation{Max Planck Institute for the Physics of Complex Systems, N\"{o}thnitzer Stra{\ss}e 38, 01187 Dresden, Germany}
\author{Zhaoyu Han}
\altaffiliation{zhan@fas.harvard.edu}
\affiliation{Department of Physics, Harvard University, Cambridge, Massachusetts 02138, USA}
\author{Biao Lian}
\altaffiliation{biao@princeton.edu}
\affiliation{Department of Physics, Princeton University, Princeton, New Jersey 08544, USA}

\begin{abstract}
We study the phase diagram of a one-dimensional spin quantum breakdown model, which has an exponential $U(1)$ symmetry with charge unit decaying as $2^{-j}$ with site position $j$. By exact diagonalization and density matrix renormalization group, we show that the model with spin $S\ge2$ exhibits an exponential $U(1)$ spontaneous symmetry-breaking (SSB) phase dubbed a quantum breakdown condensate. It exhibits a bulk gap violating the Goldstone theorem, and an edge mode only on the left edge if in open boundary condition. In a length $L$ lattice, the condensate has $\mathcal{O}(2^L)$ number of SSB ground states originating from the $\mathcal{O}(2^L)$ number of exponential $U(1)$ charge sectors, leading to a finite entropy density $\ln 2$. This enforces a first-order SSB phase transition into this phase, as observed numerically and verified in the large $S$ limit on an exactly solvable Rokhsar-Kivelson line. The condensate has an SSB order parameter being the local in-plane spin, which points in angles related by the chaotic Bernoulli (dyadic) map and thus is effectively random. Moreover, we show the condensate exhibits nondecaying local autocorrelations, and does not have an off-diagonal long-range order. The quantum breakdown condensate thus behaves as a disorder-free quantum glass and is beyond the existing classifications of phases of matter. 
\end{abstract}
\maketitle

Symmetries and spontaneous symmetry breaking (SSB) play a unifying role in the classification and characterization of quantum phases of matter. Recently, generalized symmetries~\cite{annurev:/content/journals/10.1146/annurev-conmatphys-040721-021029,LUO20241} have attracted extensive interests. On lattices, a class of on-site symmetries is the modulated symmetry \cite{Sala2022Dynamics} with symmetry charge $\hat {Q}=\sum_j f_j\hat{n}_j$, where the unit of charge $f_j$ depends on site $j$, and $\hat{n}_j$ is the particle number. Examples include dipole \cite{bergholtz2008,nakamura2012,Lake2022dipolar,schulz2019stark,Khemani2020localization,sala2020,Gorantla2022dipole,Ye2020fractonic,Ye2021fractonic}, multipole \cite{lake2022multipole,Ye2025fractonic}, subsystem symmetries \cite{chamon2005,haah2011,You2018subsystem,Burnell2022anomaly,Devakul2018classification,distler2022spontaneously,myerson2022a,myerson2022,Sfairopoulos_2023}, and more recently exponential symmetries \cite{lian2023quantum,chen2024quantum,liu20232d,hu2025bosonic,Kaplan_2016,hu_watanabe_2023,watanabe2023ground,delfino2023,Han2024topological,Sala2024exotic,wang2025exponentiallyslowthermalization1d}.  For instance, in one dimension (1D), the $k$th multipole $U(1)$ symmetry has $f_j=j^k$, and the exponential $U(1)$ symmetry with base $q\neq 1$ has $f_j=f_0q^{-j}$, for which translationally invariant models can be written down \cite{chen2026translationally}. 

Particularly, as found in a recent mean-field study \cite{hu2025bosonic}, the SSB ground state of a boson quantum breakdown model with exponential $U(1)$ symmetry, dubbed the quantum breakdown condensate, exhibits a gapped excitation spectrum violating the Goldstone theorem. This motivates us to uncover more unconventional characters of this state, and to explore its realization in more generic models.

In this Letter, we propose and study the SSB of a 1D spin quantum breakdown model [\cref{eq:spin_ham}], which has an exponential $U(1)$ symmetry with base $q=2$. Compared to the boson model~\cite{hu2025bosonic}, the spin model has a finite on-site Hilbert space and is more experimentally feasible. Employing exact diagonalization (ED) and density matrix renormalization group (DMRG), we find the model with spin $S\ge 2$ generically undergoes a phase transition from a symmetric paramagnetic (PM) phase to the exponential $U(1)$ SSB quantum breakdown condensate. We verify that the condensate has a bulk gap as found in \cite{hu2025bosonic}, and we further reveal an edge mode only existing on the left edge which alters the ED spectrum for open boundary condition. For a lattice of length $L$, we find the condensate has $\mathcal{O}(2^L)$ SSB ground states originating from $\mathcal{O}(2^L)$ number of charge sectors, which span a finite total bandwidth, in contrast to typical SSB. Also, both \cite{hu2025bosonic} and our work find the SSB phase transition being first order. We further identify a Rokhsar-Kivelson line in the large $S$ limit to be an exactly solvable first-order phase boundary. We argue that this SSB phase transition has to be first order, as enforced by a finite entropy density $\ln 2$ from $\mathcal{O}(2^L)$ ground states in the condensate. Moreover, we show the condensate has a local SSB order parameter giving in-plane spins pointing in angles related by the chaotic Bernoulli (dyadic) map \cite{renyi1957} from site to site, which typically look random. We also find it exhibits non-decaying local autocorrelations, and has no off-diagonal long range order. These facts indicate the quantum breakdown condensate behaves as a disorder-free quantum glass, and lies beyond any known classifications of phases of matter.

\begin{figure}
    \centering
    \includegraphics[width=8cm]{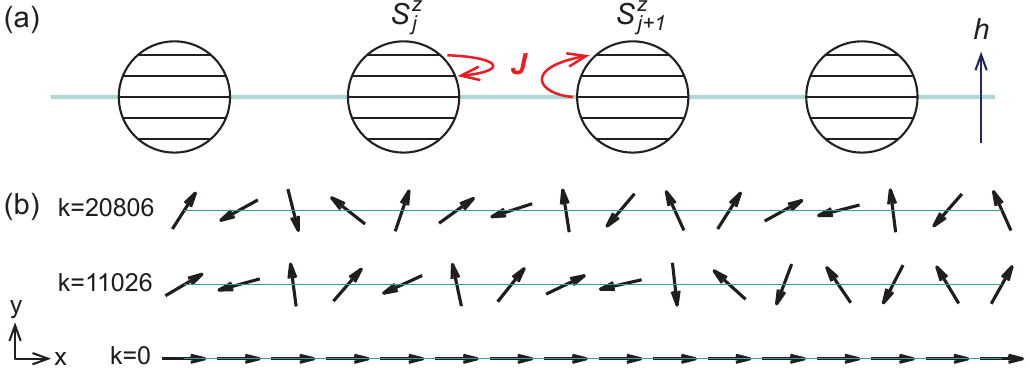}
    \caption{(a) Illustration of the spin breakdown model in Eq. \eqref{eq:spin_ham}.  (b) The in-plane spin directions on different sites for the quantum breakdown condensate under PBC with order parameter $\langle \hat S_j^+\rangle=\alpha_S e^{i2^{L-j}\theta}$. We take $L=16$ and $\theta=\frac{2\pi}{2^L-1}k$ with representative $k$ values labeled in the plot. }
    \label{fig:model}
\end{figure}

\emph{The model---}In a spin chain of length $L$ with a spin $S\ge 1$ at each site, we define the 1D spin quantum breakdown model Hamiltonian [\cref{fig:model}(a)],
\begin{equation}
\begin{split}
\hat {H}=&-\frac{J}{(2 S)^{3 / 2}}\sum_{j=1}^{L_{P\backslash O}} \left[\hat {S}_{ j}^{-}(\hat {S}_{ j+1}^+)^2 +\hat {S}_{ j}^{+}(\hat {S}_{ j+1}^{-})^2 \right]\\
&-h  \sum_{j=1}^L\left[\lambda \hat {S}_{j}^z+(1-\lambda)\frac{4 S^2}{\pi^2}\sin \left(\frac{\pi}{2S} \hat {S}_{ j}^z\right)\right]\ ,
\end{split}
\label{eq:spin_ham}
\end{equation}
where $\hat {S}_j^\nu$ ($\nu=x,y,z$) are the spin operators on site $j$, and $\hat {S}_j^\pm=\hat {S}_j^x\pm i\hat {S}_j^y$. The first term with coefficient $-J/(2S)^{3/2}$ is the inversion asymmetric breakdown interaction \cite{lian2023quantum}, in which the sum is up to $L_O=L-1$ for the open boundary condition (OBC), and up to $L_P=L$ for the periodic boundary condition (PBC) with site $L+j$ identified with site $j$. The second term is a nonlinear coupling to a $z$-direction magnetic field $h$ with $\lambda\in[0,1]$, which reduces to linear when $\lambda=1$. 

The model in \cref{eq:spin_ham} has an exponential symmetry $\hat{U}(\varphi)=\exp(i\varphi\sum_{j=1}^L 2^{L-j}\hat {S}_j^z)$, which transforms the spin on site $j$ as $\hat{U}(\varphi)(\hat {S}_j^z,\hat {S}_j^\pm) \hat{U}(\varphi)^\dagger=(\hat {S}_j^z,e^{\pm i 2^{L-j}\varphi}\hat {S}_j^\pm)$. For OBC, any angle $\varphi\in[0,2\pi)$ is allowed, which gives a $U(1)$ exponential symmetry. For PBC, the transformation on site $j=L$ should be identical to $j=0$, which enforces $(2^L-1)\varphi=0\ \mathrm{mod}\ (2\pi)$, leading to a reduction from $U(1)$ to $\mathbb{Z}_{2^L-1}$ exponential symmetry depending on system size $L$. We can rewrite $\hat{U}(\varphi)=e^{i\varphi\hat{Q}}$ as generated by an exponential charge $\hat{Q}$,
\begin{equation}\label{eq:exp_charge}
\hat{Q}=\begin{cases}
&\sum_{j=1}^L 2^{L-j}\hat {S}_j^z\ ,\qquad\qquad\qquad\quad\ (\mathrm{OBC})\\
&\sum_{j=1}^L 2^{L-j}\hat {S}_j^z\ \ \mathrm{mod}\ (2^L-1)\ .\quad (\mathrm{PBC})
\end{cases}
\end{equation}
The number of charge sectors is $N_Q=2S(2^L-1)+1$ for OBC, and $N_Q=2^L-1$ for PBC, respectively. 
As $L\rightarrow\infty$, we shall simply call $\hat{U}(\varphi)$ an exponential U(1) symmetry regardless of boundary conditions. For PBC, there is additionally a translation symmetry $\hat{T}$ sending site $j$ to $j+1$, which does not commute with the exponential symmetry: $\hat{T}\hat{U}(\varphi)\hat{T}^{-1}=\hat{U}(2\varphi)$.

\emph{Large spin limit and Rokhsar-Kivelson line---}We first analyze the phases of the model in the large spin limit. For $h>0$,  by a Holstein-Primakoff transformation $\hat{S}_z^j=S-\hat{n}_j$, $\hat{S}_j^+=(\hat{S}_j^-)^\dagger=\sqrt{2S-\hat{n}_j}\hat{a}_j$ with boson operators $\hat{a}_j, \hat{a}_j^\dagger$ and $\hat{n}_j=\hat{a}_j^\dagger \hat{a}_j$, and taking $S\rightarrow\infty$, one can map the spin model in \cref{eq:spin_ham} to the boson quantum breakdown model \cite{hu2025bosonic},
\begin{equation}\label{eq:boson-ham}
\begin{split}
\hat{H}=&-J\sum_{j=1}^{L_{P\backslash O}}\Big[\hat{a}_j (\hat{a}_{j+1}^\dagger)^2+\hat{a}_j ^\dagger(\hat{a}_{j+1})^2\Big]\\
&+\sum_{j=1}^L\Big[-\mu \hat{n}_j+\frac{U}{2}\hat{n}_j (\hat{n}_j-1)\Big]\ ,
\end{split}
\end{equation}
with a Hubbard interaction $U=(1-\lambda)h$ and chemical potential $\mu=-(1+\lambda)h/2$. As numerically revealed in \cite{hu2025bosonic}, as $J/U$ increases, the ground state of boson model in \cref{eq:boson-ham} undergoes a phase transition from a symmetric \emph{Mott insulator} to a \emph{quantum breakdown condensate} with the exponential symmetry spontaneously broken.

Here we present an analytical understanding for this SSB near a Rokhsar-Kivelson (RK) line, which is defined by $\mu U=-2J^2$ with $U>0$. On this RK line, by defining a real parameter $\alpha = 2J/U$, we can rewrite the boson Hamiltonian $\hat H$ in \cref{eq:boson-ham} (exactly for PBC, and up to boundary terms for OBC) into a semipositive definite form,
\begin{equation}\label{eq:RK_boson}
\hat{H}_{\mathrm{RK}}=\frac{U}{2}\sum_{j} \hat{O}_j^\dagger \hat{O}_j\ ,\qquad \hat{O}_j= \hat{a}_{j+1}^2 - \alpha \hat{a}_j\ .
\end{equation}
Apparently, a state $|\Psi\rangle$ satisfying $\hat{O}_j|\Psi\rangle=0$ for all $j$ gives an exact ground state of $\hat H_{\mathrm{RK}}$ with energy $E_\mathrm{gs}=0$, which has two families of solutions (unnormalized),
\begin{equation}\label{eq:boson-GS}
|\Psi_\mathrm{sym}\rangle=|0\rangle,\quad |\Psi_\text{SSB}(\theta)\rangle=\prod_{j=1}^L \exp\big(\alpha e^{i2^{L-j}\theta}\hat{a}_j^\dagger\big)|0\rangle,
\end{equation}
where $\theta\in2\pi\mathbb{R}/\mathbb{Z}$ for OBC, and $\theta\in \frac{2\pi}{2^L-1}\mathbb{Z}_{2^L-1}$ for PBC, while $|0\rangle$ is the boson vacuum. Thus, $\hat H_{\mathrm{RK}}$ in \cref{eq:RK_boson} is a frustration-free Hamiltonian with solvable ground states in \cref{eq:boson-GS} \cite{rokhsar1988,han2025frustration}. $|\Psi_\mathrm{sym}\rangle$ is a symmetric Mott insulator state with zero boson per site. The coherent state $|\Psi_\text{SSB}(\theta)\rangle$ gives an SSB quantum breakdown condensate, which has a local order parameter $\langle \hat{a}_j\rangle=\alpha e^{i2^{L-j}\theta}$ breaking the exponential $U(1)$ or $\mathbb{Z}_{2^L-1}$ symmetry \cite{hu2025bosonic}. The coherent states $|\Psi_\text{SSB}(\theta)\rangle$ can be further superposed into the charge $Q$ eigenbasis, which gives exactly one ground state $|\Psi_Q\rangle$ per charge  $Q$ sector.

Similar to the RK model \cite{rokhsar1988,han2025frustration}, the fact that both $|\Psi_\mathrm{sym}\rangle$ and $|\Psi_\text{SSB}(\theta)\rangle$ are ground states of $\hat H_{\mathrm{RK}}$ implies that the RK line $\mu U=-2J^2$ is the phase boundary of the zero-temperature SSB transition. When $J$ is perturbatively increased (decreased) away from the RK line, the state $|\Psi_\text{SSB}(\theta)\rangle$ ($|\Psi_\mathrm{sym}\rangle$) will have a lower energy, leading to the SSB (symmetric) phase. This is an energy level crossing transition with the order parameter $|\langle \hat{a}_j\rangle|$ jumping discontinuously from $0$ to $|\alpha|$, which indicates that the RK line is a \emph{first-order} phase boundary. This quantitatively matches the numerical results in \cite{hu2025bosonic}. In fact, we will argue later that the SSB phase transition of exponential $U(1)$ symmetry is generically \emph{first order}. 

\begin{figure*}[t]
    \centering
    \includegraphics[width=17cm]{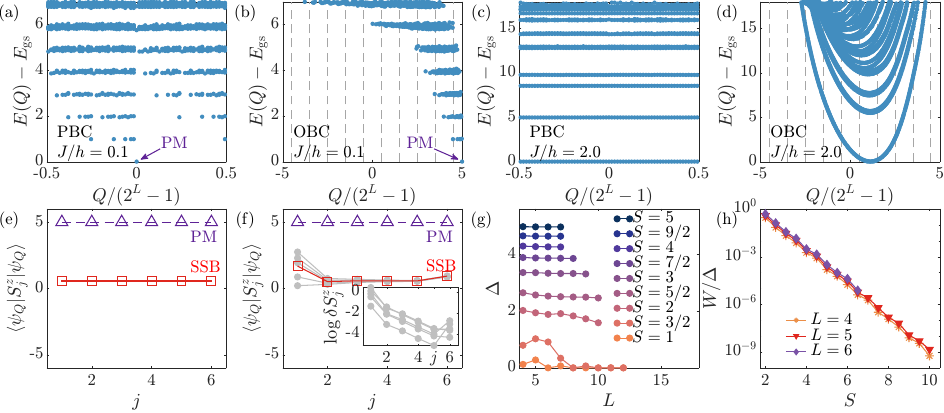}
    \caption{ED calculations for $\lambda=h=1$. (a)-(d) The ED low-energy spectra (subtracting the global ground-state energy $E_\mathrm{gs}$) versus charge sector $Q$ for $S=5$, $L=6$, in the PM phase ($J=0.1$) in (a), (b) and the SSB quantum breakdown condensate phase ($J=2$) in (c), (d). (a) and (c) are calculated with PBC; (b) and (d) are calculated with OBC. (e), (f) The expectation $\braket{\hat{S}_j^z}$ versus site $j$ for (e) PBC and (f) OBC ground states in (a)-(d). The triangular markers in (e),(f) represent the PM ground state. The square markers in (e) plot the PBC SSB ground states in all $Q$ sectors, the differences among which are negligible. The square markers in (f) are for the global SSB ground state $|\psi_{Q_0}\rangle$ which is in the $Q_0=71$ sector, while dotted markers in (f) are for SSB ground states $|\psi_{Q}\rangle$ in several other charge sectors $Q\in\{19,42,80,93,110\}$, and the (f) inset shows $\delta S_j^z\equiv\braket{\psi_Q|\hat{S}_j^z|\psi_Q}-\braket{\psi_{Q_0}|\hat{S}_j^z|\psi_{Q_0}}$ which decay exponentially in $j$. (g) The excitation gap $\Delta$ and (h) bandwidth $W$ of the SSB  ground-state manifold with PBC for different $S$ and $L$, where we fix $J=2$ and $h=\lambda=1$.
    }
    \label{fig2}
\end{figure*}

In the spin picture, the SSB order parameter $\langle \hat{a}_j\rangle$ becomes $\langle \hat {S}_j^+\rangle=\alpha_S e^{i2^{L-j}\theta}$, with $\alpha_S\neq0$. Interestingly, this points the in-plane spin components $(\langle \hat {S}_j^x\rangle,\langle \hat {S}_j^y\rangle)$ on site $j$ in an angle $\theta_j=2^{L-j}\theta\ (\mathrm{mod}\ 2\pi)$, which is typically a random looking glasslike configuration for generic $\theta$ [\cref{fig:model}(b)]. In fact, the angles follow a recursion relation $\theta_j=2\theta_{j+1}\ (\mathrm{mod}\ 2\pi)$, which is exactly the ergodic and chaotic Bernoulli (dyadic) map \cite{renyi1957} describing the Lorenz strange attractor \cite{lorenz1963,guckenheimer1979structural,tucker1999lorenz}. The SSB state is thus analogous to a spin glass, with more evidence to be shown later.

\emph{Finite spin---}For finite spin $S$, we study the Hamiltonian in \cref{eq:spin_ham} employing  ED below and DMRG in the End Matter. We find the SSB phase transition of exponential $U(1)$ symmetry robustly occurs for $S\ge 2$, with characters qualitatively agreeing with the above large $S$ limit analysis  (see End Matter and Supplemental Material \cite{supp}). Hereafter, we set $\lambda=1$, $h=1$, and focus on the example of $S=5$, for which the ground-state SSB phase transition happens at $J=J_0\approx 0.52$. 

For $J<J_0$, the model is in the symmetric phase, denoted as the \emph{paramagnetic} (PM) phase in the spin picture.  Figures \ref{fig2}(a) and \ref{fig2}(b) show the ED low-energy spectra at $J=0.1$ for $L=6$ with PBC and OBC, respectively, with the horizontal axis labeling the charge sector. The PBC and OBC spectra both have a unique gapped PM ground state, and are similar by noting that the PBC charge sectors can be viewed as the OBC charge sectors folded into the range $Q/(2^{L}-1)\in [-1/2,1/2]$ [folding boundaries marked by dashed lines in Fig.~\ref{fig2}(b)]. As shown by the triangle marker line in Figs. \ref{fig2}(e) and (f), the PM ground state has spin $\langle \hat{S}_j^z\rangle=S$ maximally polarized. 

In contrast, the ED spectra for $J>J_0$ in the SSB quantum breakdown condensate phase are notably different between PBC and OBC, as shown in Figs. \ref{fig2}(c) and (d) calculated at $J=2$ for $L=6$. For PBC, each charge $Q$ sector contributes a ground state, the energy of which we denote as $E_\mathrm{gs}(Q)$. These $2^L-1$ states $E_\mathrm{gs}(Q)$ form a nearly degenerate SSB ground-state manifold spanning an energy bandwidth $W=E_{2^L-1}-E_1$, and are separated with higher-energy states by a finite gap $\Delta=E_{2^L}-E_{2^L-1}$, where $E_n$ is the $n$th global energy level in ascending order. Figure \ref{fig2}(g) shows the gap $\Delta$ is stably finite as $L$ increases for $S\ge2$ in the SSB regime. Figure \ref{fig2}(h) implies that $W$ decays exponentially in $S$, but is nearly $L$ independent. These finite-size ED results suggest that $W$ is finite in the thermodynamic limit, which remains to be verified more rigorously. For OBC, as \cref{fig2}(d) shows, while each charge sector $Q$ still has a gap (roughly equal to the PBC gap $\Delta$), the full spectrum is gapless. The ground-state energy $E_\mathrm{gs}(Q)$ of charge sector $Q$ is well fitted by $E_\mathrm{gs}(Q)-E_\mathrm{gs}(Q_0)\approx D\left(\frac{Q-Q_0}{2^L}\right)^2$ with $D$ independent of $L$, where $Q_0$ is the charge sector of the global ground state.

To understand the unconventional ED energy spectra of the SSB quantum breakdown condensate, we first note that it has no gapless Goldstone modes as shown by \cite{hu2025bosonic}. To see this, consider perturbing the SSB order parameter into $\langle \hat{S}_j^+\rangle=\alpha_S e^{i2^{L-j}\theta}e^{i\delta\theta_j}$ with a small phase $\delta\theta_j$, which yields a Goldstone boson Lagrangian $\mathcal{L}=\frac{g_0}{2}\sum_j[(\partial_t\delta\theta_j)^2-m_b^2(\delta\theta_j-2\delta\theta_{j+1})^2]$ as enforced by the symmetry. In the bulk, it can be rewritten as $\mathcal{L}=\frac{g_0}{2}\sum_j[(\partial_t\delta\theta_j)^2-2m_b^2(\delta\theta_j-\delta\theta_{j+1})^2-m_b^2\delta\theta_j^2]\approx \frac{g_0}{2}\int dx [(\partial_t\delta\theta)^2-v_b^2(\partial_x\delta\theta)^2-m_b^2\delta\theta^2]$, which gives a bulk gap $m_b$ matching well with the ED gap $\Delta$ in \cref{fig2}(g). For OBC, there is an extra left-edge boson zero mode with $\delta\theta_j=2^{-j}\delta\theta_\mathrm{edge}$ \cite{hu2025bosonic}, which simply generates the exponential $U(1)$ transformation $e^{i2^{-L}\delta\theta_\mathrm{edge}\hat{Q}}$. Upon quantizing $\delta\theta_\mathrm{edge}$, this gives a canonical commutation relation $[\delta\hat{\theta}_\mathrm{edge},2^{-L}\hat{Q}]=i$, which indicates $\partial_t\delta\hat{\theta}_\mathrm{edge}\propto \frac{\hat{Q}-Q_0}{2^L}$ with some number $Q_0$. The Lagrangian $\mathcal{L}$ then yields an edge mode Hamiltonian $\mathcal{H}_\mathrm{edge}\propto\frac{g_0}{2}(\partial_t\delta\hat{\theta}_\mathrm{edge})^2\propto\left(\frac{\hat{Q}-Q_0}{2^L}\right)^2$, which exactly accounts for the ground-state energy difference $E_\mathrm{gs}(Q)-E_\mathrm{gs}(Q_0)$ in the OBC spectrum in \cref{fig2}(d). 
This edge mode can be further seen in Figs. \ref{fig2}(e) and (f). For PBC shown in \cref{fig2}(e), the ground state $\langle \hat{S}_j^z\rangle$ of every charge sector is spatially uniform (the line with square markers). For OBC shown in \cref{fig2}(f), the ground state $\langle \hat{S}_j^z\rangle$ of generic charge $Q$ sectors (lines with solid circle markers) differs from that of the charge $Q_0$ sector (square markers) near site $1$, and their differences $\delta S_j^z$ decay exponentially with site $j$ (\cref{fig2}(f) inset), indicating an edge mode (see detailed analysis in Supplemental Material \cite{supp}).  Thus, for OBC, the bulk spectrum is still gapped. We note that the sensitivity of energy spectrum and $\delta S_j^z$ to boundary conditions is interestingly akin to the skin effect of non-Hermitian  systems \cite{ Yao2018edge,Bergholtz2021exceptional,lin2023topological}. 

\begin{figure}[t]
    \centering
    \includegraphics[width=8.5cm]{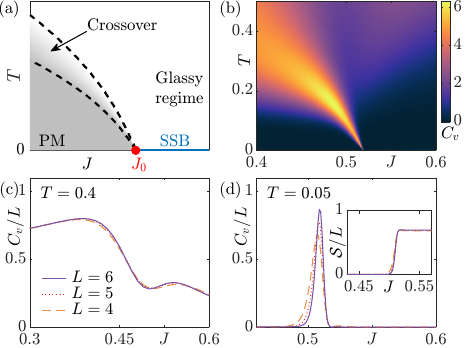}
    \caption{ (a) The phase diagram for Gibbs states. (b) The heat capacity $C_v$ for $\lambda=h=1$, $S=5$, $L=6$ with PBC. (c), (d) The specific heat $C_v/L$ versus $J$ at temperature (c) $T=0.4$ and (d) $T=0.05$, for parameters in (b). The inset of (d) shows the entropy density $\mathcal{S}/L$ at $T=0.05$. The legend in (c) labels $L$ for (c), (d).}
    \label{fig3}
\end{figure}
\emph{Phase diagram---}Since the SSB quantum breakdown condensate at $J>J_0$ has a bulk gap $\Delta>0$ above the ground-state manifold, the Mermin-Wagner theorem \cite{Mermin1966mermin, Hohenberg1967existence} does not apply, and the SSB is stable in 1D at temperature $T=0$ \cite{hu2025bosonic}. 

We now argue that the $T=0$ phase boundary between the PM phase and SSB phase is generically first order. The key observation is that the SSB phase at $J>J_0$ has $\mathcal{O}(2^L)$ number of ($2^L-1$ for PBC) nearly degenerate ground states distributed in a finite energy window, which gives a finite entropy density $\mathcal{S}/L=\ln 2$ at zero temperature, violating the third law of thermodynamics. Intrinsically, this is due to the existence of $N_Q\simeq\mathcal{O}(2^L)$ exponential U(1) charge sectors (the number of SSB ground states is generically polynomial in the number of charge sectors \cite{tasaki2019long,koma1994symmetry}). In contrast, the PM phase has a unique gapped ground state and thus an entropy density $\mathcal{S}/L=0$ at zero temperature. Explicitly, upon ignoring the energy levels above the bulk gap, we first take the $L\rightarrow\infty$ limit at temperature $T>0$, which yields a thermal entropy density regardless of the boundary condition (see Supplemental Material \cite{supp}),
\begin{equation}\label{eq:crossover}
\frac{\mathcal{S}}{L}=
\begin{cases}
&0\ ,\quad\ (J<J_T) \\
&\ln 2\ ,\ (J\ge J_T)
\end{cases} ,\ \ \  J_T=J_0-v_\mathrm{gs}^{-1}T\ln2\ ,
\end{equation}
with some $v_\text{gs}>0$. We then take the $T\rightarrow0$ limit, which reveals that the entropy density $\mathcal{S}/L$ jumps discontinuously by $\ln2$ at $J=J_0$. This enforces the $T=0$ phase boundary between PM and SSB to be \emph{first order}. This is further verified by the level crossings numerically observed in the End Matter \cref{SFigure_DMRG} and  Supplemental Material \cite{supp}. 

For 1D Gibbs states (with finite local Hilbert space) at temperatures $T>0$, it is known that fluctuations forbid any phase transition and SSB \cite{ruelle1969statistical}. Nevertheless, \cref{eq:crossover} still implies a sharp crossover at $J=J_T$ at low temperatures $T>0$, as illustrated by the \cref{fig3}(a) phase diagram. Figure \ref{fig3}(b) shows the color map of heat capacity $C_v=T(\partial \mathcal{S}/\partial T)$ from ED for $L=6$, which is indeed peaked around $J=J_T$. The specific heat $C_v/L$ as a function of $J$ is smooth at high $T$ [\cref{fig3}(c)], and shows a peak at low $T$ [\cref{fig3}(d)]. The inset of \cref{fig3}(d) further shows that the entropy density $\mathcal{S}/L$ increases sharply by approximately $\ln 2$ (sharper for the larger $L$ calculated). Note that, at $J<J_0$, the system crosses over from a more symmetric low-entropy regime to a less symmetric high-entropy regime as $T$ increases, realizing a \emph{Pomeranchuk effect} \cite{pomeranchuk1950theory,richardson1997}. 

\begin{figure}[t]
    \centering
    \includegraphics[width=8.5cm]{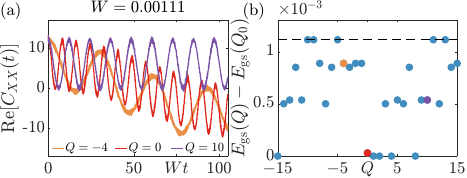}
    \caption{(a) Autocorrelation $C_{XX}(t)=\braket{\psi_Q|\hat{S}_L^x(t)\hat{S}_L^x(0)|\psi_Q}$ for PBC ground state $\ket{\psi_Q}$ of three representative charge sectors $Q$, where $J=2$, $\lambda=h=1$, $S=5$, and $L=5$. These states are highlighted in (b), which plots the energy levels in the PBC ground-state manifold (the dashed line indicates the bandwidth $W=0.00111$).}
    \label{fig4}
\end{figure}

\emph{Glassy features---}The fact that the quantum breakdown condensate possesses exponentially many low-energy eigenstates and finite entropy density $\mathcal{S}/L=\ln 2$ suggests its analogy to a glass state. A glass state is hallmarked by exponentially many equilibria and slow equilibration after a local perturbation, which are characters typically associated with quenched disorders. To examine whether the quantum breakdown condensate fulfills the defining features of a glass, we calculate the autocorrelation function $C_{XX}(t)=\langle\psi_Q|\hat{S}_j^x(t)\hat{S}_j^x(0)|\psi_Q\rangle$ of local operator $\hat{S}_j^x$ for the charge $Q$ sector ground state $|\psi_Q\rangle$ at $J>J_0$. Figure \ref{fig4} plots $C_{XX}(t)$ at $j=L$ for PBC in randomly selected charge sectors, all of which show nondecaying oscillations with periods on the order of $2\pi/W$, indicating glassy quantum dynamics (see also the End Matter). This is due to the fact that all local operators have a highly restricted connectivity only among the ground states of $\mathcal{O}(1)$ number of charge sectors--for instance, $\hat S_j^x$ only hops from charge sectors $Q$ to $Q\pm 2^{L-j}$.

Moreover, there is no off-diagonal long-range order (ODLRO) \cite{CNYang1962} two-point function associated with the exponential $U(1)$ SSB (see proof in the End Matter). The simplest symmetric two-point function formed by local order parameters is $\langle \hat{S}_j^+(\hat{S}_{j+l}^-)^{2^l}\rangle$, which vanishes identically as $l\rightarrow\infty$, since $(\hat{S}_{j+l}^-)^{2^l}=0$ for $2^l>2S$. One can go beyond the conventional description to consider a string correlator such as $\langle\hat{S}_j^+(\prod_{k=1}^{l-1}\hat{S}_{j+k}^-)(\hat{S}_{j+l}^-)^{2}\rangle$, which, however, always scales (grows or decays) exponentially in $l$ \cite{hu2025bosonic}. Therefore, the order parameter $\langle \hat{S}_j^+\rangle$ is effectively short-range correlated and is rather disordered, as depicted in \cref{fig:model}(b), making the exponential $U(1)$ SSB drastically different from the SSB of conventional (including multipole) symmetries. These observations suggest that the quantum breakdown condensate can be viewed as a quantum glass, which interestingly arises from a disorder-free quantum Hamiltonian, in contrast to most known glass models. We note that a few disorder-free quantum models with glassy dynamics exist in higher dimensions~\cite{placke2024topologicalquantumspinglass,PhysRevLett.123.040601,Ben-Ami2025}.

\emph{Discussion---}The quantum breakdown condensate from SSB of the exponential $U(1)$ symmetry is exotic in its $e^{cL}$ ($c>0$) number of SSB ground states in an $\mathcal{O}(1)$ energy window from $\mathcal{O}(e^{cL})$ number of charge sectors, and its absence of gapless Goldstone mode and ODLRO. This is different from the SSB of any other known symmetries, which at most have $\mathcal{O}(L^p)$ ($p\ge 0$) number of charge sectors. Note that the exponential $\mathbb{Z}_N$ symmetry ($N$ independent of $L$) is very different, which has at most $N$ SSB ground states \cite{hu_watanabe_2023,watanabe2023ground,delfino2023,Han2024topological}. The quantum breakdown condensate thus acts as a disorder-free quantum glass, which raises questions on its equilibration to Gibbs ensembles and its classification as a quantum phase of matter.

The exponentially many charge sectors closely resemble the Hilbert space fragmentation \cite{Khemani2020localization,sala2020,moudgalya2022thermalization,Moudgalya2022Hilbert,Moudgalya_2022}. Each charge sector has no further fragmentation except for rare cases in PBC (see  Supplemental Material \cite{supp}). Moreover, a local operator can only sparsely connect an $\mathcal{O}(1)$ number of charge sectors, leading to nondecaying autocorrelations (\cref{fig4}). This provides robustness against a class of errors in light of potentially applying the quantum breakdown condensate to store and process quantum information. In PBC, the charge of the $2^L-1$ SSB ground states naturally maps to bit strings, making it possible to store information employing distinct ground states far separated in Hamming distance, for which the damage from a local perturbation is highly constrained and correctable. It would be interesting to explore experimental realizations of the spin quantum breakdown model in platforms such as Rydberg atoms.

\begin{acknowledgments}
\emph{Acknowledgments}---We are especially grateful to David Huse for insightful discussions on phase transitions. We also thank Bo-Ting Chen, Meng Zeng, Steven Kivelson and Xiao-Liang Qi for helpful conversations. Z.~H. is supported by a Simons Investigator award, the Simons Collaboration on Ultra-Quantum Matter, which is a grant from the Simons Foundation (651440). B.~L. is supported by the National Science Foundation through Princeton University’s Materials Research Science and Engineering Center DMR-2011750, and the National Science Foundation under award DMR-2141966. Additional support is provided by the Gordon and Betty Moore Foundation through Grant GBMF8685 towards the Princeton theory program.  

\end{acknowledgments}


\begin{figure*}[t]
    \centering
    \includegraphics[width=\linewidth]{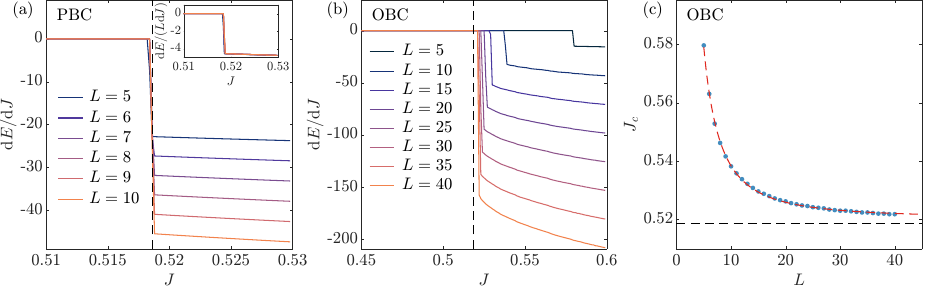}
    \caption{The derivative $\frac{\mathrm{d}E}{\mathrm{d}J}$ of the ground-state energy $E$ (obtained by DMRG) with respect to the breakdown interaction $J$ for PBC in (a) and OBC in (b). We set $h=\lambda=1$ and $S=5$ for different system sizes $L$. The inset of (a) shows that $\frac{\mathrm{d}E}{L\mathrm{d}J}$ collapse to the same curve. The dashed black line marks $J_c=0.5186$, which is the PBC phase boundary with negligible finite-size effects.  (c) The finite-size OBC phase boundary $J_c$ identified from the discontinuous jumps in (b). The red dashed line shows a fitting formula $J_c\approx 0.91L^{-1.70}+0.52$, which approaches the PBC phase boundary $J_c=0.5186$.
    } 
    \label{SFigure_DMRG}
\end{figure*}

\begin{figure}[t]
    \centering
    \includegraphics[width=8cm]{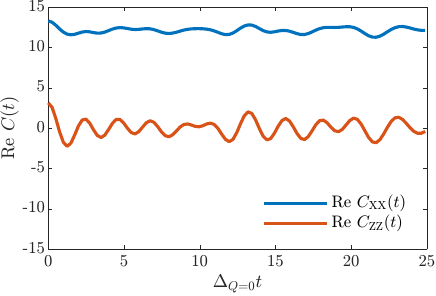}
    \caption{The $L=10$ DMRG results of the PBC autocorrelation functions $C_{XX}(t)$ and $ C_{ZZ}(t)$ with the initial state $\ket{\psi_Q}$ being the ground state in the PBC charge sector $Q=0$. We set $S=5$, $J=2$, and $h=\lambda=1$. $\Delta_{Q=0}\approx    4.9774$ is the energy gap between the ground state and first excited state in the $Q=0$ charge sector. 
    }
    \label{sfig:dmrg_evolution}
\end{figure}

\begin{figure}[t]
    \centering
    \includegraphics[width=8cm]{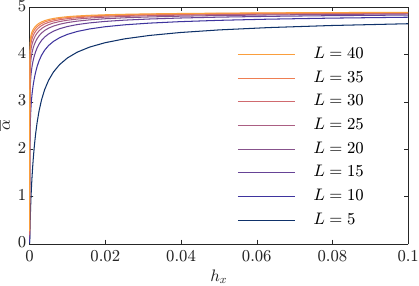}
\caption{The average order parameter $\overline{\alpha}$ [Eq. \eqref{seq:alpha}] for different system sizes $L$ and perturbations $h_x$ under OBC, obtained by large-scale DMRG.  The other parameters are set to $S=5$, $J=2$, and $h=\lambda=1$.}    \label{SFigure_dmrg_perturbation}
\end{figure}

\section*{End Matter}
In this End Matter, we use ITensor \cite{ITensor} to perform density matrix renormalization group simulations for larger system sizes, providing large-scale evidence for the zero-temperature first-order phase transition, slow relaxation dynamics, and spontaneous breaking of the exponential symmetry. In all DMRG-related simulations, we set the bond dimension $\chi=500$, which already provides converged results. We also prove that there is no ODLRO characterizing the SSB in our model.

\emph{DMRG for the ground-state phase transition---}Focusing on $S=5$, we compute the ground-state energy $E$ under both periodic and open boundary conditions and analyze its derivative with respect to the breakdown interaction $J$, namely $\mathrm{d}E/\mathrm{d}J$, for larger system sizes, as shown in Fig.~\ref{SFigure_DMRG}. Under PBC, the energy derivatives for different system sizes exhibit a clear discontinuous jump, while the inset shows a good collapse of $\mathrm{d}E/(L\mathrm{d}J)$ [Fig.~\ref{SFigure_DMRG}(a)]. These results identify a first-order transition at $J_c\simeq0.5186$ with negligible finite-size effects. On the other hand, the OBC transition point extracted from the discontinuous jump strongly depends on system size [Fig.~\ref{SFigure_DMRG}(b)], but systematically approaches the PBC value as $L$ increases. As shown in Fig.~\ref{SFigure_DMRG}(c), the finite-size scaling up to $L=40$ is well fitted by $J_c\simeq0.91L^{-1.70}+0.52$, converging to the PBC critical point $J_c\simeq0.5186$ in the thermodynamic limit. These large-scale DMRG results provide strong evidence for the existence of the first-order phase transition in large systems, with OBC finite-size effects well controlled and smoothly connected to the bulk transition identified under PBC.

\emph{PBC autocorrelation functions---}We now discuss the dynamics of the PBC autocorrelation functions on site $L$ (all sites are equivalent for PBC),
\begin{equation}
    \begin{split}
        &C_{ZZ}(t)=\braket{\psi_Q|\hat S_L^z(t)\hat S_L^z(0)|\psi_Q}
-\braket{\hat S_L^z}_Q^2\ , \\
&C_{XX}(t)=\braket{\psi_Q|\hat S_L^x(t)\hat S_L^x(0)|\psi_Q}\ ,
    \end{split}
\end{equation}
where $\ket{\psi_Q}$ is the ground state in the PBC charge sector $Q$, and $\braket{\hat S_L^z}_Q=\braket{\psi_Q|\hat S_L^z(t)|\psi_Q}$ is independent of $t$. Also, $\braket{\hat S_L^x(t)}_Q=\braket{\psi_Q|\hat S_L^x(t)|\psi_Q}\equiv0$. We denote the $n$th  ($n\ge1$, from low to high) eigenvalue and eigenstate of the Hamiltonian in charge sector $Q$ as $E_{n}^{(Q)}$ and $\ket{\phi_{n}^{(Q)}}$, so that $\ket{\psi_Q}\equiv\ket{\phi_{1}^{(Q)}}$ is the charge sector $Q$ ground state. Accordingly, $C_{XX}(t)=\sum_{Q^\prime}\sum_{n_{Q^\prime}}e^{i(E^{(Q)}_{1}-E^{(Q^\prime)}_{n_{Q^\prime}})t}|\braket{\phi^{(Q^\prime)}_{n_{Q^\prime}}|\hat{S}_L^x|\psi_Q}|^2$, where $Q^\prime$ is the label of the charge sectors that are connected to the sector $Q$ via $\hat{S}_L^x$. As shown in Fig. \ref{fig2}(c), in the SSB phase under PBC, all the excited states ($n_{Q^\prime}\ge2$) have $E_{n_{Q^\prime}}^{(Q^\prime)}-E_1^{(Q)}\ge \Delta$ and contribute fast oscillations. In contrast, all the ground states have $|E_1^{(Q)}-E_{1}^{(Q^\prime)}|\le W$, and have $\braket{\psi_{Q'}|\hat{S}_L^x|\psi_Q}\neq 0$ only if $Q'=Q\pm1$. This leads to a slow nondecaying oscillation of timescale $\tau_X\propto 1/W$, as shown in \cref{fig4}(a) calculated by ED.

As for $C_{ZZ}(t)$, we note that $\hat{S}_L^z$ acting on $\ket{\psi_Q}$ does not change the charge sector $Q$. This gives $C_{ZZ}(t)=\sum_{n_{Q}\ge 2}e^{i(E^{(Q)}_{1}-E^{(Q)}_{n_{Q}})t}|\braket{\phi^{(Q)}_{n_{Q}}|\hat{S}_L^z|\psi_Q}|^2$. Since $E^{(Q)}_{n_{Q}\ge2}-E^{(Q)}_{1}>\Delta$, we expect $C_{ZZ}(t)$ to have only a fast dynamics of timescale $\tau_Z\propto 1/\Delta$. The ED results of $C_{ZZ}(t)$ for small system sizes can be found in Supplemental Material \cite{supp}. 

We further examine the autocorrelation functions in larger system sizes by employing the charge-conserving DMRG. For instance, for the $Q=0$ sector ground state $|\psi_{Q=0}\rangle$, we perform the time-dependent variational principle to simulate the real-time evolution. As shown in Fig.~\ref{sfig:dmrg_evolution}, we find $C_{ZZ}(t)$ rapidly oscillating around zero with a timescale $\tau_Z\sim1/\Delta_{Q=0}$, where $\Delta_{Q=0}$ is the $\mathcal{O}(1)$ excitation gap in the sector $Q=0$. In contrast, $C_{XX}(t)$ shown in Fig.~\ref{sfig:dmrg_evolution} stays around a finite value within the accessible time window. The expected oscillation timescale for $C_{XX}(t)$ is $\tau_X\gtrsim10^3$, which far exceeds the accessible time window.

\emph{Symmetry-breaking perturbations---}To demonstrate SSB in the thermodynamic limit, we numerically investigate the following Hamiltonian with an exponential U(1) symmetry-breaking perturbation $h_x>0$:
\begin{equation}\label{seq:pert_Ham}
\begin{split}
  \hat  H_{\text{pert}}=\hat H-h_x\sum_{j=1}^L\hat S_j^x\ ,
    \end{split}
\end{equation}
where $\hat H$ is the quantum breakdown Hamiltonian in Eq. \eqref{eq:spin_ham}. 
Taking OBC, we employ the DMRG to find the global ground state $\ket{\Psi(h_x)}$ of $\hat H_{\text{pert}}$ and calculate the spin order parameter $\braket{\Psi(h_x)|\hat S_j^+|\Psi(h_x)}\equiv \alpha_je^{i\theta_j}$ on site $j$,  where $\alpha_j$ is the amplitude and $\theta_j$ is the phase angle.  The perturbation $h_x>0$ pins a specific symmetry-breaking direction with all $\theta_j=0$. We define the average magnitude of order parameter
\begin{equation}\label{seq:alpha}
    \overline{\alpha}\equiv \frac{1}{L}\sum_{j=1}^{L}\alpha_j,
\end{equation}
the value of which from large-scale DMRG for different system sizes $L$ are shown in Fig. \ref{SFigure_dmrg_perturbation}. The numerical results clearly indicate that
\begin{equation}
    \lim_{h_x\to 0^+}\lim_{L\to\infty}\overline{\alpha}\neq 0.
\end{equation}
Therefore, the SSB of exponential U(1) symmetry in the thermodynamic limit can be detected in the response to an infinitesimal symmetry-breaking field. Additional ED numerical results for symmetry-breaking perturbations are presented in  Supplemental Material \cite{supp}.

\emph{Proof of the absence of ODLRO---}We prove that there are no ODLRO two-point functions of local operators characterizing the exponential U(1) SSB in our spin model. Assume there exists an ODLRO two-point function $\langle \hat A_j \hat B_{j+l}\rangle$ nonvanishing as $|l|\rightarrow\infty$, where $\hat A_j$ and $\hat B_{j+l}$ are two local operators supported in the intervals of sites $[j-w,j+w]$ and $[j+l-w,j+l+w]$, respectively, with $w$ finite. As a valid ODLRO, one requires the two operators to carry nonzero exponential U(1) charges $q_{AB}$ and $-q_{AB}$, respectively, i.e., $[\hat Q,\hat A_j]=q_{AB} \hat A_j$ and $[\hat Q,\hat B_{j+l}]=-q_{AB} \hat B_{j+l}$. In a spin $S$ model, the on-site Hilbert space dimension $d=2S+1$ is finite. Accordingly, the smallest (largest) exponential U(1) charge that can be changed by an operator on site $j$ is $2^{L-j}$ ($2^{L-j}d$). Therefore, the nonzero charge $q_{AB}$ carried by operator $A_j$ must satisfy
\begin{equation}\label{eq-ODLRO-A}
2^{L-j-w}\le |q_{AB}|\le \sum_{k=-w}^w 2^{L-j-k}d<2^{L-j+w+1}d.
\end{equation}
Similarly, the nonzero charge $-q_{AB}$ carried by operator $B_{j+l}$ must satisfy
\begin{equation}\label{eq-ODLRO-B}
2^{L-j-l-w}\le |q_{AB}|<2^{L-j-l+w+1}d.
\end{equation}
As $|l|\rightarrow\infty$, Eqs. \ref{eq-ODLRO-A} and \ref{eq-ODLRO-B} necessarily contradict each other. Thus, such an ODLRO cannot exist.

\bibliography{spin_ref}

\begin{thebibliography}{65}%
\makeatletter
\providecommand \@ifxundefined [1]{%
 \@ifx{#1\undefined}
}%
\providecommand \@ifnum [1]{%
 \ifnum #1\expandafter \@firstoftwo
 \else \expandafter \@secondoftwo
 \fi
}%
\providecommand \@ifx [1]{%
 \ifx #1\expandafter \@firstoftwo
 \else \expandafter \@secondoftwo
 \fi
}%
\providecommand \natexlab [1]{#1}%
\providecommand \enquote  [1]{``#1''}%
\providecommand \bibnamefont  [1]{#1}%
\providecommand \bibfnamefont [1]{#1}%
\providecommand \citenamefont [1]{#1}%
\providecommand \href@noop [0]{\@secondoftwo}%
\providecommand \href [0]{\begingroup \@sanitize@url \@href}%
\providecommand \@href[1]{\@@startlink{#1}\@@href}%
\providecommand \@@href[1]{\endgroup#1\@@endlink}%
\providecommand \@sanitize@url [0]{\catcode `\\12\catcode `\$12\catcode
  `\&12\catcode `\#12\catcode `\^12\catcode `\_12\catcode `\%12\relax}%
\providecommand \@@startlink[1]{}%
\providecommand \@@endlink[0]{}%
\providecommand \url  [0]{\begingroup\@sanitize@url \@url }%
\providecommand \@url [1]{\endgroup\@href {#1}{\urlprefix }}%
\providecommand \urlprefix  [0]{URL }%
\providecommand \Eprint [0]{\href }%
\providecommand \doibase [0]{https://doi.org/}%
\providecommand \selectlanguage [0]{\@gobble}%
\providecommand \bibinfo  [0]{\@secondoftwo}%
\providecommand \bibfield  [0]{\@secondoftwo}%
\providecommand \translation [1]{[#1]}%
\providecommand \BibitemOpen [0]{}%
\providecommand \bibitemStop [0]{}%
\providecommand \bibitemNoStop [0]{.\EOS\space}%
\providecommand \EOS [0]{\spacefactor3000\relax}%
\providecommand \BibitemShut  [1]{\csname bibitem#1\endcsname}%
\let\auto@bib@innerbib\@empty
\bibitem [{\citenamefont
  {McGreevy}(2023)}]{annurev:/content/journals/10.1146/annurev-conmatphys-040721-021029}%
  \BibitemOpen
  \bibfield  {author} {\bibinfo {author} {\bibfnamefont {J.}~\bibnamefont
  {McGreevy}},\ }\bibfield  {title} {\bibinfo {title} {Generalized symmetries
  in condensed matter},\ }\href
  {https://doi.org/https://doi.org/10.1146/annurev-conmatphys-040721-021029}
  {\bibfield  {journal} {\bibinfo  {journal} {Annual Review of Condensed Matter
  Physics}\ }\textbf {\bibinfo {volume} {14}},\ \bibinfo {pages} {57} (\bibinfo
  {year} {2023})}\BibitemShut {NoStop}%
\bibitem [{\citenamefont {Luo}\ \emph {et~al.}(2024)\citenamefont {Luo},
  \citenamefont {Wang},\ and\ \citenamefont {Wang}}]{LUO20241}%
  \BibitemOpen
  \bibfield  {author} {\bibinfo {author} {\bibfnamefont {R.}~\bibnamefont
  {Luo}}, \bibinfo {author} {\bibfnamefont {Q.-R.}\ \bibnamefont {Wang}},\ and\
  \bibinfo {author} {\bibfnamefont {Y.-N.}\ \bibnamefont {Wang}},\ }\bibfield
  {title} {\bibinfo {title} {Lecture notes on generalized symmetries and
  applications},\ }\href
  {https://doi.org/https://doi.org/10.1016/j.physrep.2024.02.002} {\bibfield
  {journal} {\bibinfo  {journal} {Physics Reports}\ }\textbf {\bibinfo {volume}
  {1065}},\ \bibinfo {pages} {1} (\bibinfo {year} {2024})},\ \bibinfo {note}
  {lecture notes on generalized symmetries and applications}\BibitemShut
  {NoStop}%
\bibitem [{\citenamefont {Sala}\ \emph {et~al.}(2022)\citenamefont {Sala},
  \citenamefont {Lehmann}, \citenamefont {Rakovszky},\ and\ \citenamefont
  {Pollmann}}]{Sala2022Dynamics}%
  \BibitemOpen
  \bibfield  {author} {\bibinfo {author} {\bibfnamefont {P.}~\bibnamefont
  {Sala}}, \bibinfo {author} {\bibfnamefont {J.}~\bibnamefont {Lehmann}},
  \bibinfo {author} {\bibfnamefont {T.}~\bibnamefont {Rakovszky}},\ and\
  \bibinfo {author} {\bibfnamefont {F.}~\bibnamefont {Pollmann}},\ }\bibfield
  {title} {\bibinfo {title} {Dynamics in systems with modulated symmetries},\
  }\href {https://doi.org/10.1103/PhysRevLett.129.170601} {\bibfield  {journal}
  {\bibinfo  {journal} {Phys. Rev. Lett.}\ }\textbf {\bibinfo {volume} {129}},\
  \bibinfo {pages} {170601} (\bibinfo {year} {2022})}\BibitemShut {NoStop}%
\bibitem [{\citenamefont {Bergholtz}\ and\ \citenamefont
  {Karlhede}(2008)}]{bergholtz2008}%
  \BibitemOpen
  \bibfield  {author} {\bibinfo {author} {\bibfnamefont {E.~J.}\ \bibnamefont
  {Bergholtz}}\ and\ \bibinfo {author} {\bibfnamefont {A.}~\bibnamefont
  {Karlhede}},\ }\bibfield  {title} {\bibinfo {title} {Quantum hall system in
  tao-thouless limit},\ }\href {https://doi.org/10.1103/PhysRevB.77.155308}
  {\bibfield  {journal} {\bibinfo  {journal} {Phys. Rev. B}\ }\textbf {\bibinfo
  {volume} {77}},\ \bibinfo {pages} {155308} (\bibinfo {year}
  {2008})}\BibitemShut {NoStop}%
\bibitem [{\citenamefont {Nakamura}\ \emph {et~al.}(2012)\citenamefont
  {Nakamura}, \citenamefont {Wang},\ and\ \citenamefont
  {Bergholtz}}]{nakamura2012}%
  \BibitemOpen
  \bibfield  {author} {\bibinfo {author} {\bibfnamefont {M.}~\bibnamefont
  {Nakamura}}, \bibinfo {author} {\bibfnamefont {Z.-Y.}\ \bibnamefont {Wang}},\
  and\ \bibinfo {author} {\bibfnamefont {E.~J.}\ \bibnamefont {Bergholtz}},\
  }\bibfield  {title} {\bibinfo {title} {Exactly solvable fermion chain
  describing a $\ensuremath{\nu}=1/3$ fractional quantum hall state},\ }\href
  {https://doi.org/10.1103/PhysRevLett.109.016401} {\bibfield  {journal}
  {\bibinfo  {journal} {Phys. Rev. Lett.}\ }\textbf {\bibinfo {volume} {109}},\
  \bibinfo {pages} {016401} (\bibinfo {year} {2012})}\BibitemShut {NoStop}%
\bibitem [{\citenamefont {Lake}\ \emph {et~al.}(2022)\citenamefont {Lake},
  \citenamefont {Hermele},\ and\ \citenamefont {Senthil}}]{Lake2022dipolar}%
  \BibitemOpen
  \bibfield  {author} {\bibinfo {author} {\bibfnamefont {E.}~\bibnamefont
  {Lake}}, \bibinfo {author} {\bibfnamefont {M.}~\bibnamefont {Hermele}},\ and\
  \bibinfo {author} {\bibfnamefont {T.}~\bibnamefont {Senthil}},\ }\bibfield
  {title} {\bibinfo {title} {Dipolar bose-hubbard model},\ }\href
  {https://doi.org/10.1103/PhysRevB.106.064511} {\bibfield  {journal} {\bibinfo
   {journal} {Phys. Rev. B}\ }\textbf {\bibinfo {volume} {106}},\ \bibinfo
  {pages} {064511} (\bibinfo {year} {2022})}\BibitemShut {NoStop}%
\bibitem [{\citenamefont {Schulz}\ \emph {et~al.}(2019)\citenamefont {Schulz},
  \citenamefont {Hooley}, \citenamefont {Moessner},\ and\ \citenamefont
  {Pollmann}}]{schulz2019stark}%
  \BibitemOpen
  \bibfield  {author} {\bibinfo {author} {\bibfnamefont {M.}~\bibnamefont
  {Schulz}}, \bibinfo {author} {\bibfnamefont {C.~A.}\ \bibnamefont {Hooley}},
  \bibinfo {author} {\bibfnamefont {R.}~\bibnamefont {Moessner}},\ and\
  \bibinfo {author} {\bibfnamefont {F.}~\bibnamefont {Pollmann}},\ }\bibfield
  {title} {\bibinfo {title} {Stark many-body localization},\ }\href
  {https://doi.org/10.1103/PhysRevLett.122.040606} {\bibfield  {journal}
  {\bibinfo  {journal} {Phys. Rev. Lett.}\ }\textbf {\bibinfo {volume} {122}},\
  \bibinfo {pages} {040606} (\bibinfo {year} {2019})}\BibitemShut {NoStop}%
\bibitem [{\citenamefont {Khemani}\ \emph {et~al.}(2020)\citenamefont
  {Khemani}, \citenamefont {Hermele},\ and\ \citenamefont
  {Nandkishore}}]{Khemani2020localization}%
  \BibitemOpen
  \bibfield  {author} {\bibinfo {author} {\bibfnamefont {V.}~\bibnamefont
  {Khemani}}, \bibinfo {author} {\bibfnamefont {M.}~\bibnamefont {Hermele}},\
  and\ \bibinfo {author} {\bibfnamefont {R.}~\bibnamefont {Nandkishore}},\
  }\bibfield  {title} {\bibinfo {title} {Localization from hilbert space
  shattering: From theory to physical realizations},\ }\href
  {https://doi.org/10.1103/PhysRevB.101.174204} {\bibfield  {journal} {\bibinfo
   {journal} {Phys. Rev. B}\ }\textbf {\bibinfo {volume} {101}},\ \bibinfo
  {pages} {174204} (\bibinfo {year} {2020})}\BibitemShut {NoStop}%
\bibitem [{\citenamefont {Sala}\ \emph {et~al.}(2020)\citenamefont {Sala},
  \citenamefont {Rakovszky}, \citenamefont {Verresen}, \citenamefont {Knap},\
  and\ \citenamefont {Pollmann}}]{sala2020}%
  \BibitemOpen
  \bibfield  {author} {\bibinfo {author} {\bibfnamefont {P.}~\bibnamefont
  {Sala}}, \bibinfo {author} {\bibfnamefont {T.}~\bibnamefont {Rakovszky}},
  \bibinfo {author} {\bibfnamefont {R.}~\bibnamefont {Verresen}}, \bibinfo
  {author} {\bibfnamefont {M.}~\bibnamefont {Knap}},\ and\ \bibinfo {author}
  {\bibfnamefont {F.}~\bibnamefont {Pollmann}},\ }\bibfield  {title} {\bibinfo
  {title} {Ergodicity breaking arising from hilbert space fragmentation in
  dipole-conserving hamiltonians},\ }\href
  {https://doi.org/10.1103/PhysRevX.10.011047} {\bibfield  {journal} {\bibinfo
  {journal} {Phys. Rev. X}\ }\textbf {\bibinfo {volume} {10}},\ \bibinfo
  {pages} {011047} (\bibinfo {year} {2020})}\BibitemShut {NoStop}%
\bibitem [{\citenamefont {Gorantla}\ \emph {et~al.}(2022)\citenamefont
  {Gorantla}, \citenamefont {Lam}, \citenamefont {Seiberg},\ and\ \citenamefont
  {Shao}}]{Gorantla2022dipole}%
  \BibitemOpen
  \bibfield  {author} {\bibinfo {author} {\bibfnamefont {P.}~\bibnamefont
  {Gorantla}}, \bibinfo {author} {\bibfnamefont {H.~T.}\ \bibnamefont {Lam}},
  \bibinfo {author} {\bibfnamefont {N.}~\bibnamefont {Seiberg}},\ and\ \bibinfo
  {author} {\bibfnamefont {S.-H.}\ \bibnamefont {Shao}},\ }\bibfield  {title}
  {\bibinfo {title} {Global dipole symmetry, compact lifshitz theory, tensor
  gauge theory, and fractons},\ }\href
  {https://doi.org/10.1103/PhysRevB.106.045112} {\bibfield  {journal} {\bibinfo
   {journal} {Phys. Rev. B}\ }\textbf {\bibinfo {volume} {106}},\ \bibinfo
  {pages} {045112} (\bibinfo {year} {2022})}\BibitemShut {NoStop}%
\bibitem [{\citenamefont {Yuan}\ \emph {et~al.}(2020)\citenamefont {Yuan},
  \citenamefont {Chen},\ and\ \citenamefont {Ye}}]{Ye2020fractonic}%
  \BibitemOpen
  \bibfield  {author} {\bibinfo {author} {\bibfnamefont {J.-K.}\ \bibnamefont
  {Yuan}}, \bibinfo {author} {\bibfnamefont {S.~A.}\ \bibnamefont {Chen}},\
  and\ \bibinfo {author} {\bibfnamefont {P.}~\bibnamefont {Ye}},\ }\bibfield
  {title} {\bibinfo {title} {Fractonic superfluids},\ }\href
  {https://doi.org/10.1103/PhysRevResearch.2.023267} {\bibfield  {journal}
  {\bibinfo  {journal} {Phys. Rev. Res.}\ }\textbf {\bibinfo {volume} {2}},\
  \bibinfo {pages} {023267} (\bibinfo {year} {2020})}\BibitemShut {NoStop}%
\bibitem [{\citenamefont {Chen}\ \emph {et~al.}(2021)\citenamefont {Chen},
  \citenamefont {Yuan},\ and\ \citenamefont {Ye}}]{Ye2021fractonic}%
  \BibitemOpen
  \bibfield  {author} {\bibinfo {author} {\bibfnamefont {S.~A.}\ \bibnamefont
  {Chen}}, \bibinfo {author} {\bibfnamefont {J.-K.}\ \bibnamefont {Yuan}},\
  and\ \bibinfo {author} {\bibfnamefont {P.}~\bibnamefont {Ye}},\ }\bibfield
  {title} {\bibinfo {title} {Fractonic superfluids. ii. condensing
  subdimensional particles},\ }\href
  {https://doi.org/10.1103/PhysRevResearch.3.013226} {\bibfield  {journal}
  {\bibinfo  {journal} {Phys. Rev. Res.}\ }\textbf {\bibinfo {volume} {3}},\
  \bibinfo {pages} {013226} (\bibinfo {year} {2021})}\BibitemShut {NoStop}%
\bibitem [{\citenamefont {Stahl}\ \emph {et~al.}(2022)\citenamefont {Stahl},
  \citenamefont {Lake},\ and\ \citenamefont {Nandkishore}}]{lake2022multipole}%
  \BibitemOpen
  \bibfield  {author} {\bibinfo {author} {\bibfnamefont {C.}~\bibnamefont
  {Stahl}}, \bibinfo {author} {\bibfnamefont {E.}~\bibnamefont {Lake}},\ and\
  \bibinfo {author} {\bibfnamefont {R.}~\bibnamefont {Nandkishore}},\
  }\bibfield  {title} {\bibinfo {title} {Spontaneous breaking of multipole
  symmetries},\ }\href {https://doi.org/10.1103/PhysRevB.105.155107} {\bibfield
   {journal} {\bibinfo  {journal} {Phys. Rev. B}\ }\textbf {\bibinfo {volume}
  {105}},\ \bibinfo {pages} {155107} (\bibinfo {year} {2022})}\BibitemShut
  {NoStop}%
\bibitem [{\citenamefont {Wang}\ \emph
  {et~al.}(2025{\natexlab{a}})\citenamefont {Wang}, \citenamefont {Chen},\ and\
  \citenamefont {Ye}}]{Ye2025fractonic}%
  \BibitemOpen
  \bibfield  {author} {\bibinfo {author} {\bibfnamefont {H.-X.}\ \bibnamefont
  {Wang}}, \bibinfo {author} {\bibfnamefont {S.~A.}\ \bibnamefont {Chen}},\
  and\ \bibinfo {author} {\bibfnamefont {P.}~\bibnamefont {Ye}},\ }\bibfield
  {title} {\bibinfo {title} {Fractonic superfluids. iii. hybridizing higher
  moments},\ }\href {https://doi.org/10.1103/jvs9-6qkn} {\bibfield  {journal}
  {\bibinfo  {journal} {Phys. Rev. Res.}\ }\textbf {\bibinfo {volume} {7}},\
  \bibinfo {pages} {033118} (\bibinfo {year} {2025}{\natexlab{a}})}\BibitemShut
  {NoStop}%
\bibitem [{\citenamefont {Chamon}(2005)}]{chamon2005}%
  \BibitemOpen
  \bibfield  {author} {\bibinfo {author} {\bibfnamefont {C.}~\bibnamefont
  {Chamon}},\ }\bibfield  {title} {\bibinfo {title} {Quantum glassiness in
  strongly correlated clean systems: An example of topological
  overprotection},\ }\href {https://doi.org/10.1103/PhysRevLett.94.040402}
  {\bibfield  {journal} {\bibinfo  {journal} {Phys. Rev. Lett.}\ }\textbf
  {\bibinfo {volume} {94}},\ \bibinfo {pages} {040402} (\bibinfo {year}
  {2005})}\BibitemShut {NoStop}%
\bibitem [{\citenamefont {Haah}(2011)}]{haah2011}%
  \BibitemOpen
  \bibfield  {author} {\bibinfo {author} {\bibfnamefont {J.}~\bibnamefont
  {Haah}},\ }\bibfield  {title} {\bibinfo {title} {Local stabilizer codes in
  three dimensions without string logical operators},\ }\href
  {https://doi.org/10.1103/PhysRevA.83.042330} {\bibfield  {journal} {\bibinfo
  {journal} {Phys. Rev. A}\ }\textbf {\bibinfo {volume} {83}},\ \bibinfo
  {pages} {042330} (\bibinfo {year} {2011})}\BibitemShut {NoStop}%
\bibitem [{\citenamefont {You}\ \emph {et~al.}(2018)\citenamefont {You},
  \citenamefont {Devakul}, \citenamefont {Burnell},\ and\ \citenamefont
  {Sondhi}}]{You2018subsystem}%
  \BibitemOpen
  \bibfield  {author} {\bibinfo {author} {\bibfnamefont {Y.}~\bibnamefont
  {You}}, \bibinfo {author} {\bibfnamefont {T.}~\bibnamefont {Devakul}},
  \bibinfo {author} {\bibfnamefont {F.~J.}\ \bibnamefont {Burnell}},\ and\
  \bibinfo {author} {\bibfnamefont {S.~L.}\ \bibnamefont {Sondhi}},\ }\bibfield
   {title} {\bibinfo {title} {Subsystem symmetry protected topological order},\
  }\href {https://doi.org/10.1103/PhysRevB.98.035112} {\bibfield  {journal}
  {\bibinfo  {journal} {Phys. Rev. B}\ }\textbf {\bibinfo {volume} {98}},\
  \bibinfo {pages} {035112} (\bibinfo {year} {2018})}\BibitemShut {NoStop}%
\bibitem [{\citenamefont {Burnell}\ \emph {et~al.}(2022)\citenamefont
  {Burnell}, \citenamefont {Devakul}, \citenamefont {Gorantla}, \citenamefont
  {Lam},\ and\ \citenamefont {Shao}}]{Burnell2022anomaly}%
  \BibitemOpen
  \bibfield  {author} {\bibinfo {author} {\bibfnamefont {F.~J.}\ \bibnamefont
  {Burnell}}, \bibinfo {author} {\bibfnamefont {T.}~\bibnamefont {Devakul}},
  \bibinfo {author} {\bibfnamefont {P.}~\bibnamefont {Gorantla}}, \bibinfo
  {author} {\bibfnamefont {H.~T.}\ \bibnamefont {Lam}},\ and\ \bibinfo {author}
  {\bibfnamefont {S.-H.}\ \bibnamefont {Shao}},\ }\bibfield  {title} {\bibinfo
  {title} {Anomaly inflow for subsystem symmetries},\ }\href
  {https://doi.org/10.1103/PhysRevB.106.085113} {\bibfield  {journal} {\bibinfo
   {journal} {Phys. Rev. B}\ }\textbf {\bibinfo {volume} {106}},\ \bibinfo
  {pages} {085113} (\bibinfo {year} {2022})}\BibitemShut {NoStop}%
\bibitem [{\citenamefont {Devakul}\ \emph {et~al.}(2018)\citenamefont
  {Devakul}, \citenamefont {Williamson},\ and\ \citenamefont
  {You}}]{Devakul2018classification}%
  \BibitemOpen
  \bibfield  {author} {\bibinfo {author} {\bibfnamefont {T.}~\bibnamefont
  {Devakul}}, \bibinfo {author} {\bibfnamefont {D.~J.}\ \bibnamefont
  {Williamson}},\ and\ \bibinfo {author} {\bibfnamefont {Y.}~\bibnamefont
  {You}},\ }\bibfield  {title} {\bibinfo {title} {Classification of subsystem
  symmetry-protected topological phases},\ }\href
  {https://doi.org/10.1103/PhysRevB.98.235121} {\bibfield  {journal} {\bibinfo
  {journal} {Phys. Rev. B}\ }\textbf {\bibinfo {volume} {98}},\ \bibinfo
  {pages} {235121} (\bibinfo {year} {2018})}\BibitemShut {NoStop}%
\bibitem [{\citenamefont {Distler}\ \emph {et~al.}(2022)\citenamefont
  {Distler}, \citenamefont {Karch},\ and\ \citenamefont
  {Raz}}]{distler2022spontaneously}%
  \BibitemOpen
  \bibfield  {author} {\bibinfo {author} {\bibfnamefont {J.}~\bibnamefont
  {Distler}}, \bibinfo {author} {\bibfnamefont {A.}~\bibnamefont {Karch}},\
  and\ \bibinfo {author} {\bibfnamefont {A.}~\bibnamefont {Raz}},\ }\bibfield
  {title} {\bibinfo {title} {Spontaneously broken subsystem symmetries},\
  }\bibfield  {journal} {\bibinfo  {journal} {Journal of High Energy Physics}\
  }\textbf {\bibinfo {volume} {2022}},\ \href
  {https://doi.org/https://doi.org/10.1007/JHEP03(2022)016}
  {https://doi.org/10.1007/JHEP03(2022)016} (\bibinfo {year}
  {2022})\BibitemShut {NoStop}%
\bibitem [{\citenamefont {Myerson-Jain}\ \emph
  {et~al.}(2022{\natexlab{a}})\citenamefont {Myerson-Jain}, \citenamefont
  {Yan}, \citenamefont {Weld},\ and\ \citenamefont {Xu}}]{myerson2022a}%
  \BibitemOpen
  \bibfield  {author} {\bibinfo {author} {\bibfnamefont {N.~E.}\ \bibnamefont
  {Myerson-Jain}}, \bibinfo {author} {\bibfnamefont {S.}~\bibnamefont {Yan}},
  \bibinfo {author} {\bibfnamefont {D.}~\bibnamefont {Weld}},\ and\ \bibinfo
  {author} {\bibfnamefont {C.}~\bibnamefont {Xu}},\ }\bibfield  {title}
  {\bibinfo {title} {Construction of fractal order and phase transition with
  rydberg atoms},\ }\href {https://doi.org/10.1103/PhysRevLett.128.017601}
  {\bibfield  {journal} {\bibinfo  {journal} {Phys. Rev. Lett.}\ }\textbf
  {\bibinfo {volume} {128}},\ \bibinfo {pages} {017601} (\bibinfo {year}
  {2022}{\natexlab{a}})}\BibitemShut {NoStop}%
\bibitem [{\citenamefont {Myerson-Jain}\ \emph
  {et~al.}(2022{\natexlab{b}})\citenamefont {Myerson-Jain}, \citenamefont
  {Liu}, \citenamefont {Ji}, \citenamefont {Xu},\ and\ \citenamefont
  {Vijay}}]{myerson2022}%
  \BibitemOpen
  \bibfield  {author} {\bibinfo {author} {\bibfnamefont {N.~E.}\ \bibnamefont
  {Myerson-Jain}}, \bibinfo {author} {\bibfnamefont {S.}~\bibnamefont {Liu}},
  \bibinfo {author} {\bibfnamefont {W.}~\bibnamefont {Ji}}, \bibinfo {author}
  {\bibfnamefont {C.}~\bibnamefont {Xu}},\ and\ \bibinfo {author}
  {\bibfnamefont {S.}~\bibnamefont {Vijay}},\ }\bibfield  {title} {\bibinfo
  {title} {Pascal's triangle fractal symmetries},\ }\href
  {https://doi.org/10.1103/PhysRevLett.128.115301} {\bibfield  {journal}
  {\bibinfo  {journal} {Phys. Rev. Lett.}\ }\textbf {\bibinfo {volume} {128}},\
  \bibinfo {pages} {115301} (\bibinfo {year} {2022}{\natexlab{b}})}\BibitemShut
  {NoStop}%
\bibitem [{\citenamefont {Sfairopoulos}\ \emph {et~al.}(2023)\citenamefont
  {Sfairopoulos}, \citenamefont {Causer}, \citenamefont {Mair},\ and\
  \citenamefont {Garrahan}}]{Sfairopoulos_2023}%
  \BibitemOpen
  \bibfield  {author} {\bibinfo {author} {\bibfnamefont {K.}~\bibnamefont
  {Sfairopoulos}}, \bibinfo {author} {\bibfnamefont {L.}~\bibnamefont
  {Causer}}, \bibinfo {author} {\bibfnamefont {J.~F.}\ \bibnamefont {Mair}},\
  and\ \bibinfo {author} {\bibfnamefont {J.~P.}\ \bibnamefont {Garrahan}},\
  }\bibfield  {title} {\bibinfo {title} {Boundary conditions dependence of the
  phase transition in the quantum newman-moore model},\ }\href
  {https://doi.org/10.1103/PhysRevB.108.174107} {\bibfield  {journal} {\bibinfo
   {journal} {Phys. Rev. B}\ }\textbf {\bibinfo {volume} {108}},\ \bibinfo
  {pages} {174107} (\bibinfo {year} {2023})}\BibitemShut {NoStop}%
\bibitem [{\citenamefont {Lian}(2023)}]{lian2023quantum}%
  \BibitemOpen
  \bibfield  {author} {\bibinfo {author} {\bibfnamefont {B.}~\bibnamefont
  {Lian}},\ }\bibfield  {title} {\bibinfo {title} {Quantum breakdown model:
  From many-body localization to chaos with scars},\ }\href
  {https://doi.org/10.1103/PhysRevB.107.115171} {\bibfield  {journal} {\bibinfo
   {journal} {Phys. Rev. B}\ }\textbf {\bibinfo {volume} {107}},\ \bibinfo
  {pages} {115171} (\bibinfo {year} {2023})}\BibitemShut {NoStop}%
\bibitem [{\citenamefont {Chen}\ \emph {et~al.}(2024)\citenamefont {Chen},
  \citenamefont {Prem}, \citenamefont {Regnault},\ and\ \citenamefont
  {Lian}}]{chen2024quantum}%
  \BibitemOpen
  \bibfield  {author} {\bibinfo {author} {\bibfnamefont {B.-T.}\ \bibnamefont
  {Chen}}, \bibinfo {author} {\bibfnamefont {A.}~\bibnamefont {Prem}}, \bibinfo
  {author} {\bibfnamefont {N.}~\bibnamefont {Regnault}},\ and\ \bibinfo
  {author} {\bibfnamefont {B.}~\bibnamefont {Lian}},\ }\bibfield  {title}
  {\bibinfo {title} {Quantum fragmentation in the extended quantum breakdown
  model},\ }\bibfield  {journal} {\bibinfo  {journal} {Physical Review B}\
  }\textbf {\bibinfo {volume} {110}},\ \href
  {https://doi.org/10.1103/physrevb.110.165109} {10.1103/physrevb.110.165109}
  (\bibinfo {year} {2024})\BibitemShut {NoStop}%
\bibitem [{\citenamefont {Liu}\ and\ \citenamefont {Lian}(2025)}]{liu20232d}%
  \BibitemOpen
  \bibfield  {author} {\bibinfo {author} {\bibfnamefont {X.}~\bibnamefont
  {Liu}}\ and\ \bibinfo {author} {\bibfnamefont {B.}~\bibnamefont {Lian}},\
  }\bibfield  {title} {\bibinfo {title} {Two-dimensional quantum breakdown
  model with krylov subspace many-body localization},\ }\href
  {https://doi.org/10.1103/PhysRevB.111.054302} {\bibfield  {journal} {\bibinfo
   {journal} {Phys. Rev. B}\ }\textbf {\bibinfo {volume} {111}},\ \bibinfo
  {pages} {054302} (\bibinfo {year} {2025})}\BibitemShut {NoStop}%
\bibitem [{\citenamefont {Hu}\ and\ \citenamefont
  {Lian}(2025)}]{hu2025bosonic}%
  \BibitemOpen
  \bibfield  {author} {\bibinfo {author} {\bibfnamefont {Y.-M.}\ \bibnamefont
  {Hu}}\ and\ \bibinfo {author} {\bibfnamefont {B.}~\bibnamefont {Lian}},\
  }\bibfield  {title} {\bibinfo {title} {Bosonic quantum breakdown hubbard
  model},\ }\href {https://doi.org/10.1103/1r4m-7psy} {\bibfield  {journal}
  {\bibinfo  {journal} {Phys. Rev. B}\ }\textbf {\bibinfo {volume} {112}},\
  \bibinfo {pages} {L100504} (\bibinfo {year} {2025})}\BibitemShut {NoStop}%
\bibitem [{\citenamefont {Kaplan}\ and\ \citenamefont
  {Rattazzi}(2016)}]{Kaplan_2016}%
  \BibitemOpen
  \bibfield  {author} {\bibinfo {author} {\bibfnamefont {D.~E.}\ \bibnamefont
  {Kaplan}}\ and\ \bibinfo {author} {\bibfnamefont {R.}~\bibnamefont
  {Rattazzi}},\ }\bibfield  {title} {\bibinfo {title} {Large field excursions
  and approximate discrete symmetries from a clockwork axion},\ }\href
  {https://doi.org/10.1103/PhysRevD.93.085007} {\bibfield  {journal} {\bibinfo
  {journal} {Phys. Rev. D}\ }\textbf {\bibinfo {volume} {93}},\ \bibinfo
  {pages} {085007} (\bibinfo {year} {2016})}\BibitemShut {NoStop}%
\bibitem [{\citenamefont {Hu}\ and\ \citenamefont
  {Watanabe}(2023)}]{hu_watanabe_2023}%
  \BibitemOpen
  \bibfield  {author} {\bibinfo {author} {\bibfnamefont {Y.}~\bibnamefont
  {Hu}}\ and\ \bibinfo {author} {\bibfnamefont {H.}~\bibnamefont {Watanabe}},\
  }\bibfield  {title} {\bibinfo {title} {Spontaneous symmetry breaking without
  ground state degeneracy in generalized $n$-state clock model},\ }\href
  {https://doi.org/10.1103/PhysRevB.107.195139} {\bibfield  {journal} {\bibinfo
   {journal} {Phys. Rev. B}\ }\textbf {\bibinfo {volume} {107}},\ \bibinfo
  {pages} {195139} (\bibinfo {year} {2023})}\BibitemShut {NoStop}%
\bibitem [{\citenamefont {Watanabe}\ \emph {et~al.}(2023)\citenamefont
  {Watanabe}, \citenamefont {Cheng},\ and\ \citenamefont
  {Fuji}}]{watanabe2023ground}%
  \BibitemOpen
  \bibfield  {author} {\bibinfo {author} {\bibfnamefont {H.}~\bibnamefont
  {Watanabe}}, \bibinfo {author} {\bibfnamefont {M.}~\bibnamefont {Cheng}},\
  and\ \bibinfo {author} {\bibfnamefont {Y.}~\bibnamefont {Fuji}},\ }\bibfield
  {title} {\bibinfo {title} {Ground state degeneracy on torus in a family of zn
  toric code},\ }\href@noop {} {\bibfield  {journal} {\bibinfo  {journal}
  {Journal of Mathematical Physics}\ }\textbf {\bibinfo {volume} {64}}
  (\bibinfo {year} {2023})}\BibitemShut {NoStop}%
\bibitem [{\citenamefont {Delfino}\ \emph {et~al.}(2026)\citenamefont
  {Delfino}, \citenamefont {Chamon},\ and\ \citenamefont {You}}]{delfino2023}%
  \BibitemOpen
  \bibfield  {author} {\bibinfo {author} {\bibfnamefont {G.}~\bibnamefont
  {Delfino}}, \bibinfo {author} {\bibfnamefont {C.}~\bibnamefont {Chamon}},\
  and\ \bibinfo {author} {\bibfnamefont {Y.}~\bibnamefont {You}},\ }\bibfield
  {title} {\bibinfo {title} {Topological order and fractons from gauging
  exponential symmetries},\ }\href {https://doi.org/10.1103/vvhq-wnll}
  {\bibfield  {journal} {\bibinfo  {journal} {Phys. Rev. B}\ }\textbf {\bibinfo
  {volume} {113}},\ \bibinfo {pages} {045130} (\bibinfo {year}
  {2026})}\BibitemShut {NoStop}%
\bibitem [{\citenamefont {Han}\ \emph {et~al.}(2024)\citenamefont {Han},
  \citenamefont {Lake}, \citenamefont {Lam}, \citenamefont {Verresen},\ and\
  \citenamefont {You}}]{Han2024topological}%
  \BibitemOpen
  \bibfield  {author} {\bibinfo {author} {\bibfnamefont {J.~H.}\ \bibnamefont
  {Han}}, \bibinfo {author} {\bibfnamefont {E.}~\bibnamefont {Lake}}, \bibinfo
  {author} {\bibfnamefont {H.~T.}\ \bibnamefont {Lam}}, \bibinfo {author}
  {\bibfnamefont {R.}~\bibnamefont {Verresen}},\ and\ \bibinfo {author}
  {\bibfnamefont {Y.}~\bibnamefont {You}},\ }\bibfield  {title} {\bibinfo
  {title} {Topological quantum chains protected by dipolar and other modulated
  symmetries},\ }\href {https://doi.org/10.1103/PhysRevB.109.125121} {\bibfield
   {journal} {\bibinfo  {journal} {Phys. Rev. B}\ }\textbf {\bibinfo {volume}
  {109}},\ \bibinfo {pages} {125121} (\bibinfo {year} {2024})}\BibitemShut
  {NoStop}%
\bibitem [{\citenamefont {Sala}\ \emph {et~al.}(2024)\citenamefont {Sala},
  \citenamefont {You}, \citenamefont {Hauschild},\ and\ \citenamefont
  {Motrunich}}]{Sala2024exotic}%
  \BibitemOpen
  \bibfield  {author} {\bibinfo {author} {\bibfnamefont {P.}~\bibnamefont
  {Sala}}, \bibinfo {author} {\bibfnamefont {Y.}~\bibnamefont {You}}, \bibinfo
  {author} {\bibfnamefont {J.}~\bibnamefont {Hauschild}},\ and\ \bibinfo
  {author} {\bibfnamefont {O.}~\bibnamefont {Motrunich}},\ }\bibfield  {title}
  {\bibinfo {title} {Exotic quantum liquids in bose-hubbard models with
  spatially modulated symmetries},\ }\href
  {https://doi.org/10.1103/PhysRevB.109.014406} {\bibfield  {journal} {\bibinfo
   {journal} {Phys. Rev. B}\ }\textbf {\bibinfo {volume} {109}},\ \bibinfo
  {pages} {014406} (\bibinfo {year} {2024})}\BibitemShut {NoStop}%
\bibitem [{\citenamefont {Wang}\ \emph
  {et~al.}(2025{\natexlab{b}})\citenamefont {Wang}, \citenamefont
  {Balasubramanian}, \citenamefont {Han}, \citenamefont {Lake}, \citenamefont
  {Chen},\ and\ \citenamefont
  {Yang}}]{wang2025exponentiallyslowthermalization1d}%
  \BibitemOpen
  \bibfield  {author} {\bibinfo {author} {\bibfnamefont {C.}~\bibnamefont
  {Wang}}, \bibinfo {author} {\bibfnamefont {S.}~\bibnamefont
  {Balasubramanian}}, \bibinfo {author} {\bibfnamefont {Y.}~\bibnamefont
  {Han}}, \bibinfo {author} {\bibfnamefont {E.}~\bibnamefont {Lake}}, \bibinfo
  {author} {\bibfnamefont {X.}~\bibnamefont {Chen}},\ and\ \bibinfo {author}
  {\bibfnamefont {Z.-C.}\ \bibnamefont {Yang}},\ }\href
  {https://arxiv.org/abs/2501.13930} {\bibinfo {title} {Exponentially slow
  thermalization in 1d fragmented dynamics}} (\bibinfo {year}
  {2025}{\natexlab{b}}),\ \Eprint {https://arxiv.org/abs/2501.13930}
  {arXiv:2501.13930 [quant-ph]} \BibitemShut {NoStop}%
\bibitem [{\citenamefont {Chen}\ \emph {et~al.}(2026)\citenamefont {Chen},
  \citenamefont {Zhou},\ and\ \citenamefont {Lian}}]{chen2026translationally}%
  \BibitemOpen
  \bibfield  {author} {\bibinfo {author} {\bibfnamefont {B.-T.}\ \bibnamefont
  {Chen}}, \bibinfo {author} {\bibfnamefont {Z.}~\bibnamefont {Zhou}},\ and\
  \bibinfo {author} {\bibfnamefont {B.}~\bibnamefont {Lian}},\ }\href
  {https://arxiv.org/abs/2606.07952} {\bibinfo {title} {Translationally
  covariant modulated symmetries: Classification and goldstone}} (\bibinfo
  {year} {2026}),\ \Eprint {https://arxiv.org/abs/2606.07952} {arXiv:2606.07952
  [cond-mat.str-el]} \BibitemShut {NoStop}%
\bibitem [{\citenamefont {R{\'e}nyi}(1957)}]{renyi1957}%
  \BibitemOpen
  \bibfield  {author} {\bibinfo {author} {\bibfnamefont {A.}~\bibnamefont
  {R{\'e}nyi}},\ }\bibfield  {title} {\bibinfo {title} {Representations for
  real numbers and their ergodic properties},\ }\href@noop {} {\bibfield
  {journal} {\bibinfo  {journal} {Acta Math. Acad. Sci. Hungar}\ }\textbf
  {\bibinfo {volume} {8}},\ \bibinfo {pages} {477} (\bibinfo {year}
  {1957})}\BibitemShut {NoStop}%
\bibitem [{\citenamefont {Rokhsar}\ and\ \citenamefont
  {Kivelson}(1988)}]{rokhsar1988}%
  \BibitemOpen
  \bibfield  {author} {\bibinfo {author} {\bibfnamefont {D.~S.}\ \bibnamefont
  {Rokhsar}}\ and\ \bibinfo {author} {\bibfnamefont {S.~A.}\ \bibnamefont
  {Kivelson}},\ }\bibfield  {title} {\bibinfo {title} {Superconductivity and
  the quantum hard-core dimer gas},\ }\href
  {https://doi.org/10.1103/PhysRevLett.61.2376} {\bibfield  {journal} {\bibinfo
   {journal} {Phys. Rev. Lett.}\ }\textbf {\bibinfo {volume} {61}},\ \bibinfo
  {pages} {2376} (\bibinfo {year} {1988})}\BibitemShut {NoStop}%
\bibitem [{\citenamefont {Han}\ and\ \citenamefont
  {Kivelson}(2025)}]{han2025frustration}%
  \BibitemOpen
  \bibfield  {author} {\bibinfo {author} {\bibfnamefont {Z.}~\bibnamefont
  {Han}}\ and\ \bibinfo {author} {\bibfnamefont {S.~A.}\ \bibnamefont
  {Kivelson}},\ }\bibfield  {title} {\bibinfo {title} {Models of interacting
  bosons with exact ground states: A unified approach},\ }\href
  {https://doi.org/10.1103/PhysRevB.111.174520} {\bibfield  {journal} {\bibinfo
   {journal} {Phys. Rev. B}\ }\textbf {\bibinfo {volume} {111}},\ \bibinfo
  {pages} {174520} (\bibinfo {year} {2025})}\BibitemShut {NoStop}%
\bibitem [{\citenamefont {{Lorenz}}(1963)}]{lorenz1963}%
  \BibitemOpen
  \bibfield  {author} {\bibinfo {author} {\bibfnamefont {E.~N.}\ \bibnamefont
  {{Lorenz}}},\ }\bibfield  {title} {\bibinfo {title} {{Deterministic
  Nonperiodic Flow.}},\ }\href
  {https://doi.org/10.1175/1520-0469(1963)020<0130:DNF>2.0.CO;2} {\bibfield
  {journal} {\bibinfo  {journal} {Journal of the Atmospheric Sciences}\
  }\textbf {\bibinfo {volume} {20}},\ \bibinfo {pages} {130} (\bibinfo {year}
  {1963})}\BibitemShut {NoStop}%
\bibitem [{\citenamefont {Guckenheimer}\ and\ \citenamefont
  {Williams}(1979)}]{guckenheimer1979structural}%
  \BibitemOpen
  \bibfield  {author} {\bibinfo {author} {\bibfnamefont {J.}~\bibnamefont
  {Guckenheimer}}\ and\ \bibinfo {author} {\bibfnamefont {R.~F.}\ \bibnamefont
  {Williams}},\ }\bibfield  {title} {\bibinfo {title} {Structural stability of
  lorenz attractors},\ }\href@noop {} {\bibfield  {journal} {\bibinfo
  {journal} {Publications Math{\'e}matiques de l'IH{\'E}S}\ }\textbf {\bibinfo
  {volume} {50}},\ \bibinfo {pages} {59} (\bibinfo {year} {1979})}\BibitemShut
  {NoStop}%
\bibitem [{\citenamefont {Tucker}(1999)}]{tucker1999lorenz}%
  \BibitemOpen
  \bibfield  {author} {\bibinfo {author} {\bibfnamefont {W.}~\bibnamefont
  {Tucker}},\ }\bibfield  {title} {\bibinfo {title} {The lorenz attractor
  exists},\ }\href@noop {} {\bibfield  {journal} {\bibinfo  {journal} {Comptes
  Rendus de l'Acad{\'e}mie des Sciences-Series I-Mathematics}\ }\textbf
  {\bibinfo {volume} {328}},\ \bibinfo {pages} {1197} (\bibinfo {year}
  {1999})}\BibitemShut {NoStop}%
\bibitem [{sup()}]{supp}%
  \BibitemOpen
  \href@noop {} {}\bibinfo {note} {See Supplemental Material for details, which
  also includes references
  \cite{oganesyan2007,atas2013,haake2019quantum,Giraud2022Probing,Ji2020Categorical}.}\BibitemShut
  {Stop}%
\bibitem [{\citenamefont {Yao}\ and\ \citenamefont {Wang}(2018)}]{Yao2018edge}%
  \BibitemOpen
  \bibfield  {author} {\bibinfo {author} {\bibfnamefont {S.}~\bibnamefont
  {Yao}}\ and\ \bibinfo {author} {\bibfnamefont {Z.}~\bibnamefont {Wang}},\
  }\bibfield  {title} {\bibinfo {title} {Edge states and topological invariants
  of non-hermitian systems},\ }\href
  {https://doi.org/10.1103/PhysRevLett.121.086803} {\bibfield  {journal}
  {\bibinfo  {journal} {Phys. Rev. Lett.}\ }\textbf {\bibinfo {volume} {121}},\
  \bibinfo {pages} {086803} (\bibinfo {year} {2018})}\BibitemShut {NoStop}%
\bibitem [{\citenamefont {Bergholtz}\ \emph {et~al.}(2021)\citenamefont
  {Bergholtz}, \citenamefont {Budich},\ and\ \citenamefont
  {Kunst}}]{Bergholtz2021exceptional}%
  \BibitemOpen
  \bibfield  {author} {\bibinfo {author} {\bibfnamefont {E.~J.}\ \bibnamefont
  {Bergholtz}}, \bibinfo {author} {\bibfnamefont {J.~C.}\ \bibnamefont
  {Budich}},\ and\ \bibinfo {author} {\bibfnamefont {F.~K.}\ \bibnamefont
  {Kunst}},\ }\bibfield  {title} {\bibinfo {title} {Exceptional topology of
  non-hermitian systems},\ }\href
  {https://doi.org/10.1103/RevModPhys.93.015005} {\bibfield  {journal}
  {\bibinfo  {journal} {Rev. Mod. Phys.}\ }\textbf {\bibinfo {volume} {93}},\
  \bibinfo {pages} {015005} (\bibinfo {year} {2021})}\BibitemShut {NoStop}%
\bibitem [{\citenamefont {Lin}\ \emph {et~al.}(2023)\citenamefont {Lin},
  \citenamefont {Tai}, \citenamefont {Li},\ and\ \citenamefont
  {Lee}}]{lin2023topological}%
  \BibitemOpen
  \bibfield  {author} {\bibinfo {author} {\bibfnamefont {R.}~\bibnamefont
  {Lin}}, \bibinfo {author} {\bibfnamefont {T.}~\bibnamefont {Tai}}, \bibinfo
  {author} {\bibfnamefont {L.}~\bibnamefont {Li}},\ and\ \bibinfo {author}
  {\bibfnamefont {C.~H.}\ \bibnamefont {Lee}},\ }\bibfield  {title} {\bibinfo
  {title} {Topological non-hermitian skin effect},\ }\href@noop {} {\bibfield
  {journal} {\bibinfo  {journal} {Frontiers of Physics}\ }\textbf {\bibinfo
  {volume} {18}},\ \bibinfo {pages} {53605} (\bibinfo {year}
  {2023})}\BibitemShut {NoStop}%
\bibitem [{\citenamefont {Mermin}\ and\ \citenamefont
  {Wagner}(1966)}]{Mermin1966mermin}%
  \BibitemOpen
  \bibfield  {author} {\bibinfo {author} {\bibfnamefont {N.~D.}\ \bibnamefont
  {Mermin}}\ and\ \bibinfo {author} {\bibfnamefont {H.}~\bibnamefont
  {Wagner}},\ }\bibfield  {title} {\bibinfo {title} {Absence of ferromagnetism
  or antiferromagnetism in one- or two-dimensional isotropic heisenberg
  models},\ }\href {https://doi.org/10.1103/PhysRevLett.17.1133} {\bibfield
  {journal} {\bibinfo  {journal} {Phys. Rev. Lett.}\ }\textbf {\bibinfo
  {volume} {17}},\ \bibinfo {pages} {1133} (\bibinfo {year}
  {1966})}\BibitemShut {NoStop}%
\bibitem [{\citenamefont {Hohenberg}(1967)}]{Hohenberg1967existence}%
  \BibitemOpen
  \bibfield  {author} {\bibinfo {author} {\bibfnamefont {P.~C.}\ \bibnamefont
  {Hohenberg}},\ }\bibfield  {title} {\bibinfo {title} {Existence of long-range
  order in one and two dimensions},\ }\href
  {https://doi.org/10.1103/PhysRev.158.383} {\bibfield  {journal} {\bibinfo
  {journal} {Phys. Rev.}\ }\textbf {\bibinfo {volume} {158}},\ \bibinfo {pages}
  {383} (\bibinfo {year} {1967})}\BibitemShut {NoStop}%
\bibitem [{\citenamefont {Tasaki}(2019)}]{tasaki2019long}%
  \BibitemOpen
  \bibfield  {author} {\bibinfo {author} {\bibfnamefont {H.}~\bibnamefont
  {Tasaki}},\ }\bibfield  {title} {\bibinfo {title} {Long-range
  order,“tower” of states, and symmetry breaking in lattice quantum
  systems},\ }\href@noop {} {\bibfield  {journal} {\bibinfo  {journal} {Journal
  of Statistical Physics}\ }\textbf {\bibinfo {volume} {174}},\ \bibinfo
  {pages} {735} (\bibinfo {year} {2019})}\BibitemShut {NoStop}%
\bibitem [{\citenamefont {Koma}\ and\ \citenamefont
  {Tasaki}(1994)}]{koma1994symmetry}%
  \BibitemOpen
  \bibfield  {author} {\bibinfo {author} {\bibfnamefont {T.}~\bibnamefont
  {Koma}}\ and\ \bibinfo {author} {\bibfnamefont {H.}~\bibnamefont {Tasaki}},\
  }\bibfield  {title} {\bibinfo {title} {Symmetry breaking and finite-size
  effects in quantum many-body systems},\ }\href
  {https://doi.org/https://doi.org/10.1007/BF02188685} {\bibfield  {journal}
  {\bibinfo  {journal} {Journal of statistical physics}\ }\textbf {\bibinfo
  {volume} {76}},\ \bibinfo {pages} {745} (\bibinfo {year} {1994})}\BibitemShut
  {NoStop}%
\bibitem [{\citenamefont {Ruelle}(1969)}]{ruelle1969statistical}%
  \BibitemOpen
  \bibfield  {author} {\bibinfo {author} {\bibfnamefont {D.}~\bibnamefont
  {Ruelle}},\ }\href@noop {} {\emph {\bibinfo {title} {Statistical mechanics:
  Rigorous results}}}\ (\bibinfo  {publisher} {World Scientific},\ \bibinfo
  {year} {1969})\BibitemShut {NoStop}%
\bibitem [{\citenamefont {Pomeranchuk}(1950)}]{pomeranchuk1950theory}%
  \BibitemOpen
  \bibfield  {author} {\bibinfo {author} {\bibfnamefont {I.}~\bibnamefont
  {Pomeranchuk}},\ }\bibfield  {title} {\bibinfo {title} {On the theory of
  liquid he$^3$},\ }\href@noop {} {\bibfield  {journal} {\bibinfo  {journal}
  {Zhur. Eksptl'. i Teoret. Fiz.}\ }\textbf {\bibinfo {volume} {20}} (\bibinfo
  {year} {1950})}\BibitemShut {NoStop}%
\bibitem [{\citenamefont {Richardson}(1997)}]{richardson1997}%
  \BibitemOpen
  \bibfield  {author} {\bibinfo {author} {\bibfnamefont {R.~C.}\ \bibnamefont
  {Richardson}},\ }\bibfield  {title} {\bibinfo {title} {The pomeranchuk
  effect},\ }\href {https://doi.org/10.1103/RevModPhys.69.683} {\bibfield
  {journal} {\bibinfo  {journal} {Rev. Mod. Phys.}\ }\textbf {\bibinfo {volume}
  {69}},\ \bibinfo {pages} {683} (\bibinfo {year} {1997})}\BibitemShut
  {NoStop}%
\bibitem [{\citenamefont {Yang}(1962)}]{CNYang1962}%
  \BibitemOpen
  \bibfield  {author} {\bibinfo {author} {\bibfnamefont {C.~N.}\ \bibnamefont
  {Yang}},\ }\bibfield  {title} {\bibinfo {title} {Concept of off-diagonal
  long-range order and the quantum phases of liquid he and of
  superconductors},\ }\href {https://doi.org/10.1103/RevModPhys.34.694}
  {\bibfield  {journal} {\bibinfo  {journal} {Rev. Mod. Phys.}\ }\textbf
  {\bibinfo {volume} {34}},\ \bibinfo {pages} {694} (\bibinfo {year}
  {1962})}\BibitemShut {NoStop}%
\bibitem [{\citenamefont {Placke}\ \emph {et~al.}(2024)\citenamefont {Placke},
  \citenamefont {Rakovszky}, \citenamefont {Breuckmann},\ and\ \citenamefont
  {Khemani}}]{placke2024topologicalquantumspinglass}%
  \BibitemOpen
  \bibfield  {author} {\bibinfo {author} {\bibfnamefont {B.}~\bibnamefont
  {Placke}}, \bibinfo {author} {\bibfnamefont {T.}~\bibnamefont {Rakovszky}},
  \bibinfo {author} {\bibfnamefont {N.~P.}\ \bibnamefont {Breuckmann}},\ and\
  \bibinfo {author} {\bibfnamefont {V.}~\bibnamefont {Khemani}},\ }\href
  {https://arxiv.org/abs/2412.13248} {\bibinfo {title} {Topological quantum
  spin glass order and its realization in qldpc codes}} (\bibinfo {year}
  {2024}),\ \Eprint {https://arxiv.org/abs/2412.13248} {arXiv:2412.13248
  [quant-ph]} \BibitemShut {NoStop}%
\bibitem [{\citenamefont {Feldmeier}\ \emph {et~al.}(2019)\citenamefont
  {Feldmeier}, \citenamefont {Pollmann},\ and\ \citenamefont
  {Knap}}]{PhysRevLett.123.040601}%
  \BibitemOpen
  \bibfield  {author} {\bibinfo {author} {\bibfnamefont {J.}~\bibnamefont
  {Feldmeier}}, \bibinfo {author} {\bibfnamefont {F.}~\bibnamefont
  {Pollmann}},\ and\ \bibinfo {author} {\bibfnamefont {M.}~\bibnamefont
  {Knap}},\ }\bibfield  {title} {\bibinfo {title} {Emergent glassy dynamics in
  a quantum dimer model},\ }\href
  {https://doi.org/10.1103/PhysRevLett.123.040601} {\bibfield  {journal}
  {\bibinfo  {journal} {Phys. Rev. Lett.}\ }\textbf {\bibinfo {volume} {123}},\
  \bibinfo {pages} {040601} (\bibinfo {year} {2019})}\BibitemShut {NoStop}%
\bibitem [{\citenamefont {{Ben-Ami}}\ \emph {et~al.}(2025)\citenamefont
  {{Ben-Ami}}, \citenamefont {{Heyl}},\ and\ \citenamefont
  {{Moessner}}}]{Ben-Ami2025}%
  \BibitemOpen
  \bibfield  {author} {\bibinfo {author} {\bibfnamefont {T.}~\bibnamefont
  {{Ben-Ami}}}, \bibinfo {author} {\bibfnamefont {M.}~\bibnamefont {{Heyl}}},\
  and\ \bibinfo {author} {\bibfnamefont {R.}~\bibnamefont {{Moessner}}},\
  }\bibfield  {title} {\bibinfo {title} {{Many-body cages: disorder-free
  glassiness from flat bands in Fock space, and many-body Rabi oscillations}},\
  }\href {https://doi.org/10.48550/arXiv.2504.13086} {\bibfield  {journal}
  {\bibinfo  {journal} {arXiv e-prints}\ ,\ \bibinfo {eid} {arXiv:2504.13086}}
  (\bibinfo {year} {2025})},\ \Eprint {https://arxiv.org/abs/2504.13086}
  {arXiv:2504.13086 [cond-mat.quant-gas]} \BibitemShut {NoStop}%
\bibitem [{\citenamefont {Moudgalya}\ \emph
  {et~al.}(2022{\natexlab{a}})\citenamefont {Moudgalya}, \citenamefont {Prem},
  \citenamefont {Nandkishore}, \citenamefont {Regnault},\ and\ \citenamefont
  {Bernevig}}]{moudgalya2022thermalization}%
  \BibitemOpen
  \bibfield  {author} {\bibinfo {author} {\bibfnamefont {S.}~\bibnamefont
  {Moudgalya}}, \bibinfo {author} {\bibfnamefont {A.}~\bibnamefont {Prem}},
  \bibinfo {author} {\bibfnamefont {R.}~\bibnamefont {Nandkishore}}, \bibinfo
  {author} {\bibfnamefont {N.}~\bibnamefont {Regnault}},\ and\ \bibinfo
  {author} {\bibfnamefont {B.~A.}\ \bibnamefont {Bernevig}},\ }\bibfield
  {title} {\bibinfo {title} {Thermalization and its absence within krylov
  subspaces of a constrained hamiltonian},\ }in\ \href
  {https://doi.org/10.1142/9789811231711_0009} {\emph {\bibinfo {booktitle}
  {Memorial Volume for Shoucheng Zhang}}}\ (\bibinfo  {publisher} {World
  Scientific},\ \bibinfo {year} {2022})\ pp.\ \bibinfo {pages}
  {147--209}\BibitemShut {NoStop}%
\bibitem [{\citenamefont {Moudgalya}\ and\ \citenamefont
  {Motrunich}(2022)}]{Moudgalya2022Hilbert}%
  \BibitemOpen
  \bibfield  {author} {\bibinfo {author} {\bibfnamefont {S.}~\bibnamefont
  {Moudgalya}}\ and\ \bibinfo {author} {\bibfnamefont {O.~I.}\ \bibnamefont
  {Motrunich}},\ }\bibfield  {title} {\bibinfo {title} {Hilbert space
  fragmentation and commutant algebras},\ }\href
  {https://doi.org/10.1103/PhysRevX.12.011050} {\bibfield  {journal} {\bibinfo
  {journal} {Phys. Rev. X}\ }\textbf {\bibinfo {volume} {12}},\ \bibinfo
  {pages} {011050} (\bibinfo {year} {2022})}\BibitemShut {NoStop}%
\bibitem [{\citenamefont {Moudgalya}\ \emph
  {et~al.}(2022{\natexlab{b}})\citenamefont {Moudgalya}, \citenamefont
  {Bernevig},\ and\ \citenamefont {Regnault}}]{Moudgalya_2022}%
  \BibitemOpen
  \bibfield  {author} {\bibinfo {author} {\bibfnamefont {S.}~\bibnamefont
  {Moudgalya}}, \bibinfo {author} {\bibfnamefont {B.~A.}\ \bibnamefont
  {Bernevig}},\ and\ \bibinfo {author} {\bibfnamefont {N.}~\bibnamefont
  {Regnault}},\ }\bibfield  {title} {\bibinfo {title} {Quantum many-body scars
  and hilbert space fragmentation: a review of exact results},\ }\href
  {https://doi.org/10.1088/1361-6633/ac73a0} {\bibfield  {journal} {\bibinfo
  {journal} {Reports on Progress in Physics}\ }\textbf {\bibinfo {volume}
  {85}},\ \bibinfo {pages} {086501} (\bibinfo {year}
  {2022}{\natexlab{b}})}\BibitemShut {NoStop}%
\bibitem [{\citenamefont {Fishman}\ \emph {et~al.}(2022)\citenamefont
  {Fishman}, \citenamefont {White},\ and\ \citenamefont
  {Stoudenmire}}]{ITensor}%
  \BibitemOpen
  \bibfield  {author} {\bibinfo {author} {\bibfnamefont {M.}~\bibnamefont
  {Fishman}}, \bibinfo {author} {\bibfnamefont {S.~R.}\ \bibnamefont {White}},\
  and\ \bibinfo {author} {\bibfnamefont {E.~M.}\ \bibnamefont {Stoudenmire}},\
  }\bibfield  {title} {\bibinfo {title} {{The ITensor Software Library for
  Tensor Network Calculations}},\ }\href
  {https://doi.org/10.21468/SciPostPhysCodeb.4} {\bibfield  {journal} {\bibinfo
   {journal} {SciPost Phys. Codebases}\ ,\ \bibinfo {pages} {4}} (\bibinfo
  {year} {2022})}\BibitemShut {NoStop}%
\bibitem [{\citenamefont {Oganesyan}\ and\ \citenamefont
  {Huse}(2007)}]{oganesyan2007}%
  \BibitemOpen
  \bibfield  {author} {\bibinfo {author} {\bibfnamefont {V.}~\bibnamefont
  {Oganesyan}}\ and\ \bibinfo {author} {\bibfnamefont {D.~A.}\ \bibnamefont
  {Huse}},\ }\bibfield  {title} {\bibinfo {title} {Localization of interacting
  fermions at high temperature},\ }\href
  {https://doi.org/10.1103/PhysRevB.75.155111} {\bibfield  {journal} {\bibinfo
  {journal} {Phys. Rev. B}\ }\textbf {\bibinfo {volume} {75}},\ \bibinfo
  {pages} {155111} (\bibinfo {year} {2007})}\BibitemShut {NoStop}%
\bibitem [{\citenamefont {Atas}\ \emph {et~al.}(2013)\citenamefont {Atas},
  \citenamefont {Bogomolny}, \citenamefont {Giraud},\ and\ \citenamefont
  {Roux}}]{atas2013}%
  \BibitemOpen
  \bibfield  {author} {\bibinfo {author} {\bibfnamefont {Y.~Y.}\ \bibnamefont
  {Atas}}, \bibinfo {author} {\bibfnamefont {E.}~\bibnamefont {Bogomolny}},
  \bibinfo {author} {\bibfnamefont {O.}~\bibnamefont {Giraud}},\ and\ \bibinfo
  {author} {\bibfnamefont {G.}~\bibnamefont {Roux}},\ }\bibfield  {title}
  {\bibinfo {title} {Distribution of the ratio of consecutive level spacings in
  random matrix ensembles},\ }\href
  {https://doi.org/10.1103/PhysRevLett.110.084101} {\bibfield  {journal}
  {\bibinfo  {journal} {Phys. Rev. Lett.}\ }\textbf {\bibinfo {volume} {110}},\
  \bibinfo {pages} {084101} (\bibinfo {year} {2013})}\BibitemShut {NoStop}%
\bibitem [{\citenamefont {Haake}\ \emph {et~al.}(2019)\citenamefont {Haake},
  \citenamefont {Gnutzmann},\ and\ \citenamefont {Ku{\'s}}}]{haake2019quantum}%
  \BibitemOpen
  \bibfield  {author} {\bibinfo {author} {\bibfnamefont {F.}~\bibnamefont
  {Haake}}, \bibinfo {author} {\bibfnamefont {S.}~\bibnamefont {Gnutzmann}},\
  and\ \bibinfo {author} {\bibfnamefont {M.}~\bibnamefont {Ku{\'s}}},\ }\href
  {https://doi.org/10.1007/978-3-319-97580-1} {\emph {\bibinfo {title} {Quantum
  Signatures of Chaos}}}\ (\bibinfo  {publisher} {Springer Cham},\ \bibinfo
  {year} {2019})\BibitemShut {NoStop}%
\bibitem [{\citenamefont {Giraud}\ \emph {et~al.}(2022)\citenamefont {Giraud},
  \citenamefont {Mac\'e}, \citenamefont {Vernier},\ and\ \citenamefont
  {Alet}}]{Giraud2022Probing}%
  \BibitemOpen
  \bibfield  {author} {\bibinfo {author} {\bibfnamefont {O.}~\bibnamefont
  {Giraud}}, \bibinfo {author} {\bibfnamefont {N.}~\bibnamefont {Mac\'e}},
  \bibinfo {author} {\bibfnamefont {E.}~\bibnamefont {Vernier}},\ and\ \bibinfo
  {author} {\bibfnamefont {F.}~\bibnamefont {Alet}},\ }\bibfield  {title}
  {\bibinfo {title} {Probing symmetries of quantum many-body systems through
  gap ratio statistics},\ }\href {https://doi.org/10.1103/PhysRevX.12.011006}
  {\bibfield  {journal} {\bibinfo  {journal} {Phys. Rev. X}\ }\textbf {\bibinfo
  {volume} {12}},\ \bibinfo {pages} {011006} (\bibinfo {year}
  {2022})}\BibitemShut {NoStop}%
\bibitem [{\citenamefont {Ji}\ and\ \citenamefont
  {Wen}(2020)}]{Ji2020Categorical}%
  \BibitemOpen
  \bibfield  {author} {\bibinfo {author} {\bibfnamefont {W.}~\bibnamefont
  {Ji}}\ and\ \bibinfo {author} {\bibfnamefont {X.-G.}\ \bibnamefont {Wen}},\
  }\bibfield  {title} {\bibinfo {title} {Categorical symmetry and noninvertible
  anomaly in symmetry-breaking and topological phase transitions},\ }\href
  {https://doi.org/10.1103/PhysRevResearch.2.033417} {\bibfield  {journal}
  {\bibinfo  {journal} {Phys. Rev. Res.}\ }\textbf {\bibinfo {volume} {2}},\
  \bibinfo {pages} {033417} (\bibinfo {year} {2020})}\BibitemShut {NoStop}%
\end{thebibliography}%


\begin{thebibliography}{7}%
\makeatletter
\providecommand \@ifxundefined [1]{%
 \@ifx{#1\undefined}
}%
\providecommand \@ifnum [1]{%
 \ifnum #1\expandafter \@firstoftwo
 \else \expandafter \@secondoftwo
 \fi
}%
\providecommand \@ifx [1]{%
 \ifx #1\expandafter \@firstoftwo
 \else \expandafter \@secondoftwo
 \fi
}%
\providecommand \natexlab [1]{#1}%
\providecommand \enquote  [1]{``#1''}%
\providecommand \bibnamefont  [1]{#1}%
\providecommand \bibfnamefont [1]{#1}%
\providecommand \citenamefont [1]{#1}%
\providecommand \href@noop [0]{\@secondoftwo}%
\providecommand \href [0]{\begingroup \@sanitize@url \@href}%
\providecommand \@href[1]{\@@startlink{#1}\@@href}%
\providecommand \@@href[1]{\endgroup#1\@@endlink}%
\providecommand \@sanitize@url [0]{\catcode `\\12\catcode `\$12\catcode
  `\&12\catcode `\#12\catcode `\^12\catcode `\_12\catcode `\%12\relax}%
\providecommand \@@startlink[1]{}%
\providecommand \@@endlink[0]{}%
\providecommand \url  [0]{\begingroup\@sanitize@url \@url }%
\providecommand \@url [1]{\endgroup\@href {#1}{\urlprefix }}%
\providecommand \urlprefix  [0]{URL }%
\providecommand \Eprint [0]{\href }%
\providecommand \doibase [0]{https://doi.org/}%
\providecommand \selectlanguage [0]{\@gobble}%
\providecommand \bibinfo  [0]{\@secondoftwo}%
\providecommand \bibfield  [0]{\@secondoftwo}%
\providecommand \translation [1]{[#1]}%
\providecommand \BibitemOpen [0]{}%
\providecommand \bibitemStop [0]{}%
\providecommand \bibitemNoStop [0]{.\EOS\space}%
\providecommand \EOS [0]{\spacefactor3000\relax}%
\providecommand \BibitemShut  [1]{\csname bibitem#1\endcsname}%
\let\auto@bib@innerbib\@empty
\bibitem [{\citenamefont {Oganesyan}\ and\ \citenamefont
  {Huse}(2007)}]{oganesyan2007}%
  \BibitemOpen
  \bibfield  {author} {\bibinfo {author} {\bibfnamefont {V.}~\bibnamefont
  {Oganesyan}}\ and\ \bibinfo {author} {\bibfnamefont {D.~A.}\ \bibnamefont
  {Huse}},\ }\bibfield  {title} {\bibinfo {title} {Localization of interacting
  fermions at high temperature},\ }\href
  {https://doi.org/10.1103/PhysRevB.75.155111} {\bibfield  {journal} {\bibinfo
  {journal} {Phys. Rev. B}\ }\textbf {\bibinfo {volume} {75}},\ \bibinfo
  {pages} {155111} (\bibinfo {year} {2007})}\BibitemShut {NoStop}%
\bibitem [{\citenamefont {Atas}\ \emph {et~al.}(2013)\citenamefont {Atas},
  \citenamefont {Bogomolny}, \citenamefont {Giraud},\ and\ \citenamefont
  {Roux}}]{atas2013}%
  \BibitemOpen
  \bibfield  {author} {\bibinfo {author} {\bibfnamefont {Y.~Y.}\ \bibnamefont
  {Atas}}, \bibinfo {author} {\bibfnamefont {E.}~\bibnamefont {Bogomolny}},
  \bibinfo {author} {\bibfnamefont {O.}~\bibnamefont {Giraud}},\ and\ \bibinfo
  {author} {\bibfnamefont {G.}~\bibnamefont {Roux}},\ }\bibfield  {title}
  {\bibinfo {title} {Distribution of the ratio of consecutive level spacings in
  random matrix ensembles},\ }\href
  {https://doi.org/10.1103/PhysRevLett.110.084101} {\bibfield  {journal}
  {\bibinfo  {journal} {Phys. Rev. Lett.}\ }\textbf {\bibinfo {volume} {110}},\
  \bibinfo {pages} {084101} (\bibinfo {year} {2013})}\BibitemShut {NoStop}%
\bibitem [{\citenamefont {Haake}\ \emph {et~al.}(2019)\citenamefont {Haake},
  \citenamefont {Gnutzmann},\ and\ \citenamefont {Ku{\'s}}}]{haake2019quantum}%
  \BibitemOpen
  \bibfield  {author} {\bibinfo {author} {\bibfnamefont {F.}~\bibnamefont
  {Haake}}, \bibinfo {author} {\bibfnamefont {S.}~\bibnamefont {Gnutzmann}},\
  and\ \bibinfo {author} {\bibfnamefont {M.}~\bibnamefont {Ku{\'s}}},\ }\href
  {https://doi.org/10.1007/978-3-319-97580-1} {\emph {\bibinfo {title} {Quantum
  Signatures of Chaos}}}\ (\bibinfo  {publisher} {Springer Cham},\ \bibinfo
  {year} {2019})\BibitemShut {NoStop}%
\bibitem [{\citenamefont {Giraud}\ \emph {et~al.}(2022)\citenamefont {Giraud},
  \citenamefont {Mac\'e}, \citenamefont {Vernier},\ and\ \citenamefont
  {Alet}}]{Giraud2022Probing}%
  \BibitemOpen
  \bibfield  {author} {\bibinfo {author} {\bibfnamefont {O.}~\bibnamefont
  {Giraud}}, \bibinfo {author} {\bibfnamefont {N.}~\bibnamefont {Mac\'e}},
  \bibinfo {author} {\bibfnamefont {E.}~\bibnamefont {Vernier}},\ and\ \bibinfo
  {author} {\bibfnamefont {F.}~\bibnamefont {Alet}},\ }\bibfield  {title}
  {\bibinfo {title} {Probing symmetries of quantum many-body systems through
  gap ratio statistics},\ }\href {https://doi.org/10.1103/PhysRevX.12.011006}
  {\bibfield  {journal} {\bibinfo  {journal} {Phys. Rev. X}\ }\textbf {\bibinfo
  {volume} {12}},\ \bibinfo {pages} {011006} (\bibinfo {year}
  {2022})}\BibitemShut {NoStop}%
\bibitem [{\citenamefont {Ruelle}(1969)}]{ruelle1969statistical}%
  \BibitemOpen
  \bibfield  {author} {\bibinfo {author} {\bibfnamefont {D.}~\bibnamefont
  {Ruelle}},\ }\href@noop {} {\emph {\bibinfo {title} {Statistical mechanics:
  Rigorous results}}}\ (\bibinfo  {publisher} {World Scientific},\ \bibinfo
  {year} {1969})\BibitemShut {NoStop}%
\bibitem [{\citenamefont {Ji}\ and\ \citenamefont
  {Wen}(2020)}]{Ji2020Categorical}%
  \BibitemOpen
  \bibfield  {author} {\bibinfo {author} {\bibfnamefont {W.}~\bibnamefont
  {Ji}}\ and\ \bibinfo {author} {\bibfnamefont {X.-G.}\ \bibnamefont {Wen}},\
  }\bibfield  {title} {\bibinfo {title} {Categorical symmetry and noninvertible
  anomaly in symmetry-breaking and topological phase transitions},\ }\href
  {https://doi.org/10.1103/PhysRevResearch.2.033417} {\bibfield  {journal}
  {\bibinfo  {journal} {Phys. Rev. Res.}\ }\textbf {\bibinfo {volume} {2}},\
  \bibinfo {pages} {033417} (\bibinfo {year} {2020})}\BibitemShut {NoStop}%
\bibitem [{\citenamefont {Han}\ \emph {et~al.}(2024)\citenamefont {Han},
  \citenamefont {Lake}, \citenamefont {Lam}, \citenamefont {Verresen},\ and\
  \citenamefont {You}}]{Han2024Topological}%
  \BibitemOpen
  \bibfield  {author} {\bibinfo {author} {\bibfnamefont {J.~H.}\ \bibnamefont
  {Han}}, \bibinfo {author} {\bibfnamefont {E.}~\bibnamefont {Lake}}, \bibinfo
  {author} {\bibfnamefont {H.~T.}\ \bibnamefont {Lam}}, \bibinfo {author}
  {\bibfnamefont {R.}~\bibnamefont {Verresen}},\ and\ \bibinfo {author}
  {\bibfnamefont {Y.}~\bibnamefont {You}},\ }\bibfield  {title} {\bibinfo
  {title} {Topological quantum chains protected by dipolar and other modulated
  symmetries},\ }\href {https://doi.org/10.1103/PhysRevB.109.125121} {\bibfield
   {journal} {\bibinfo  {journal} {Phys. Rev. B}\ }\textbf {\bibinfo {volume}
  {109}},\ \bibinfo {pages} {125121} (\bibinfo {year} {2024})}\BibitemShut
  {NoStop}%
\end{thebibliography}%

\end{document}


\title{Supplemental Material: Exponential U(1) Symmetry-Breaking Phase as a Disorder-Free Quantum Glass}

\author{Yu-Min Hu}
\affiliation{Max Planck Institute for the Physics of Complex Systems, N\"{o}thnitzer Stra{\ss}e 38, 01187 Dresden, Germany}
\author{Zhaoyu Han}
\altaffiliation{zhan@fas.harvard.edu}
\affiliation{Department of Physics, Harvard University, Cambridge, Massachusetts 02138, USA}
\author{Biao Lian}
\altaffiliation{biao@princeton.edu}
\affiliation{Department of Physics, Princeton University, Princeton, New Jersey 08544, USA}

\maketitle

\tableofcontents

\section{ED energy spectra for other spins $S$}
\begin{figure}[t]
    \centering
    \includegraphics[width=\linewidth]{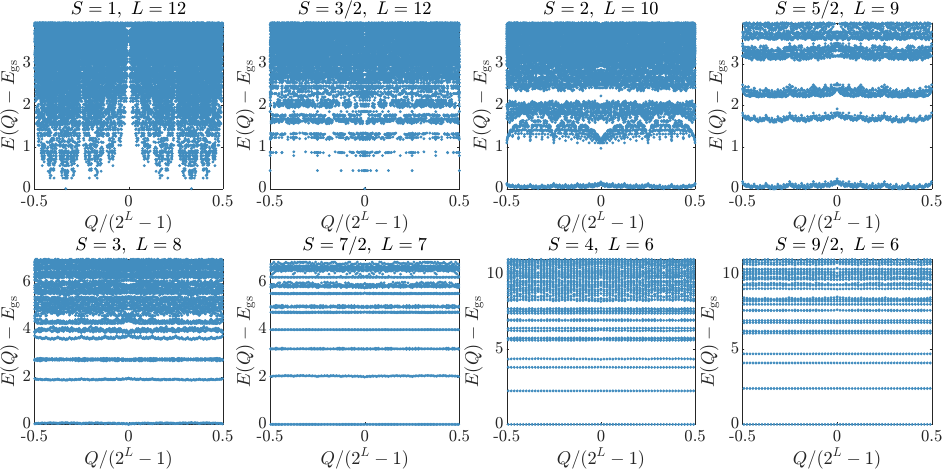}
    \caption{The PBC low-energy spectra of the model with respect to charge sector $Q$, for $J=1$ and $h=0$ (deeply in the SSB phase), for which $\hat H$ has no $\lambda$ dependence. Different choices of $S$ and $L$ are labeled in each panel.}
    \label{SFigure_ED_S}
\end{figure}

\begin{figure}[t]
    \centering
    \includegraphics[width=0.5\linewidth]{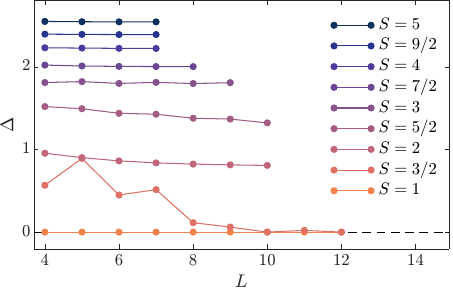}
    \caption{ The excitation gap $\Delta$ [Eq. \eqref{seq:gap}] of the SSB ground-state manifold with PBC for different $S$ and $L$, where we fix $J=1$ and $h=0$ (the same as Fig. \ref{SFigure_ED_S}), for which $\hat H$ has no $\lambda$ dependence.}
    \label{SFigure_Scaling}
\end{figure}

\begin{figure}[t]
    \centering
    \includegraphics[width=\linewidth]{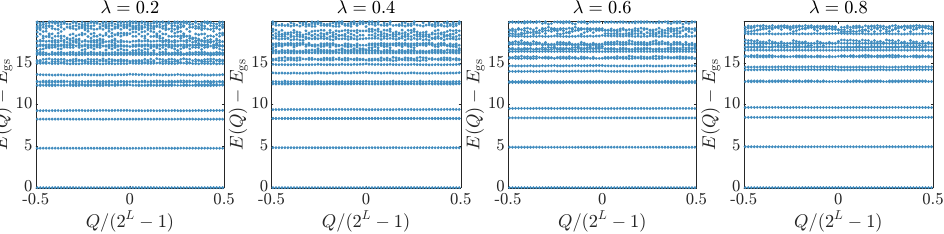}
    \caption{The PBC low-energy spectra of the model with respect to charge sector $Q$, for $S=5$, $L=6$, $J=2$, and $h=1$ (in the SSB phase). Different choices of $\lambda$ are labeled in each panel.}
    \label{SFigure_ED_Lambda}
\end{figure}
In this section, we present additional ED numerical results to show that the physics discussed in the main text generally occurs for many other spin $S$ values. For convenience, we repeat the quantum breakdown Hamiltonian [Eq.~(1) of the main text] here:
\begin{equation}
\hat {H}=-\frac{J}{(2 S)^{3 / 2}}\sum_{j=1}^{L_{P\backslash O}} \left[\hat {S}_{ j}^{-}(\hat {S}_{ j+1}^+)^2 +\hat {S}_{ j}^{+}(\hat {S}_{ j+1}^{-})^2 \right]-h  \sum_{j=1}^L\left[\lambda \hat {S}_{j}^z+(1-\lambda)\frac{4 S^2}{\pi^2}\sin \left(\frac{\pi}{2S} \hat {S}_{ j}^z\right)\right],
\label{seq:spin_ham}
\end{equation}
where $L_O=L-1$ for open boundary condition (OBC), and $L_P=L$ for periodic boundary condition (PBC) with site $L+j$ identified with site $j$.  

We first consider the PBC Hamiltonian with only the breakdown interaction, corresponding to $h=0$ in Eq.~\eqref{seq:spin_ham}, for which $\hat {H}$ has no $\lambda$ dependence (so we do not need to specify $\lambda$). This case allows us to directly identify the low-energy physics deep in the possible spontaneous-symmetry-breaking (SSB) phase (if existing), without the competition from on-site interactions. The corresponding numerical results obtained from exact diagonalization (ED) are presented in Figs. \ref{SFigure_ED_S} and \ref{SFigure_Scaling}. 

Figure~\ref{SFigure_ED_S} shows typical PBC low-energy spectra for different $S$. The spectra for $S=1$ and $S=3/2$ differ markedly from those for $S\geq2$. For $S\geq2$, a subset of eigenstates is clearly separated from the higher-energy spectrum, signaling a nearly degenerate ground-state manifold with exponentially many states. By contrast, no such gapped ground-state manifold appears for $S=1$ or $S=3/2$. This numerical observation leads to the conclusion in the main text: for $S\geq2$, the nearly degenerate ground-state manifold contains $2^L-1$ eigenstates, each being the ground state in a distinct charge sector $Q$.

To further characterize whether the SSB phase exists, we define the gap above the SSB ground-state manifold in PBC (note that the SSB phase has $2^L-1$ ground states):
\begin{equation}\label{seq:gap}
\Delta= E_{2^L}-E_{2^L-1},
\end{equation}
where $E_k$ is the $k$-th energy eigenvalue of the PBC Hamiltonian in ascending order. We examine the gap $\Delta$ as a function of system size $L$ (up to the largest size calculable by ED). As shown in Fig.~\ref{SFigure_Scaling} where we set $J=1$ and $h=0$, $\Delta$ clearly distinguishes between $S=1,3/2$ and $S\geq2$. For $S\geq2$, a finite $\Delta$ indicates a gapped ground-state manifold consisting of $2^L-1$ nearly degenerate eigenstates associated with spontaneous symmetry breaking (SSB). By contrast, the vanishing $\Delta$ for $S=1$ and $S=3/2$ implies the absence of spontaneous breaking of the exponential symmetry. Figure 2(g) of the main text shows a similar plot of $\Delta$ for $J=2$ and $h=\lambda=1$, which suggests the same conclusion.

The numerical results shown in Figs.~\ref{SFigure_ED_S} and \ref{SFigure_Scaling} demonstrate that the low-energy physics for $S=1$ and $S=3/2$ is fundamentally different from that for $S\geq 2$. The $S\ge2$ regime is characterized by the spontaneous breaking of the exponential symmetry, which is the focus of this work. A detailed analysis of the $S\le 3/2$ case is beyond the scope of this work and thus is deferred for future investigation.

Figure~\ref{SFigure_ED_Lambda} presents the PBC low-energy spectra for different $\lambda$. For sufficiently large $J/h$ (here $J/h=2$), the system enters the spontaneous-symmetry-breaking (SSB) phase, where the ground states from different PBC charge sectors become nearly degenerate. We find that the ED spectra are qualitatively similar for different $\lambda$, suggesting that the SSB physics of the exponential U(1) symmetry is independent of the specific form of symmetric on-site interactions.

\section{Level statistics in each charge sector}\label{sec:LevelSpacing}
\begin{figure}[t]
    \centering
    \includegraphics[width=\linewidth]{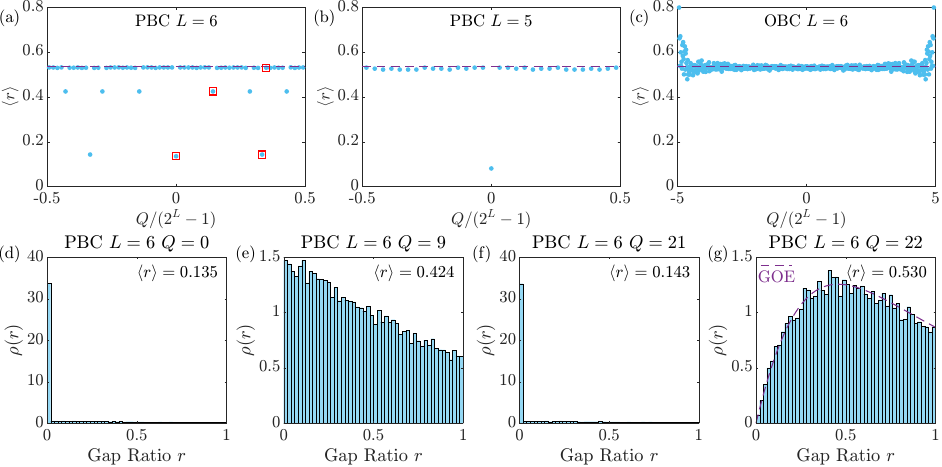}
    \caption{In all plots, we fix the parameters $S=5$, $J=2$ and $h=\lambda=1$.  (a-c) The mean values of gap ratios for different charge sectors, with the boundary conditions and the system size shown in each plot. The dashed line is the theoretical value 0.536 of the random Gaussian orthogonal ensemble.  (d-g) The gap-ratio distribution for four representative PBC charge sectors marked as red squares in (a). The dashed line in (g) represents the Gaussian orthogonal ensemble (GOE). }
    \label{SFigure_LevelStatistics}
\end{figure}
We now investigate the level spacing statistics for the spin quantum breakdown model. Without loss of generality, we focus on the parameters $S=5$ and $J=2$ and $h=\lambda=1$. These parameters are the same as in the Figs. 2(c) and 2(d) of the main text. In each exponential charge sector $Q$, we have many eigenenergies $E^{(Q)}_1\le\cdots\le E^{(Q)}_{d_Q}$  in increasing order, where $d_Q$ is the Hilbert space dimension of the charge sector $Q$ for a fixed system size. In each sector $Q$, we can calculate the gap ratio \cite{oganesyan2007}
\begin{equation}
    r_n=\frac{\min(\Delta_n,\Delta_{n+1})}{\max(\Delta_n,\Delta_{n+1})}
\end{equation}
where $\Delta_n=E_{n+1}^{(Q)}-E_n^{(Q)}$ is the gap between two neighboring energies. 

Figs. \ref{SFigure_LevelStatistics} (a-c) shows the mean gap ratio $\braket{r}=\frac{1}{d_Q-2}\sum_{n=1}^{d_Q-2}r_n$ for each $Q$ sector with different boundary conditions and system sizes  \cite{oganesyan2007,atas2013,haake2019quantum,Giraud2022Probing} . These numerical results indicate that, regardless of boundary conditions and system sizes, most charge sectors have a mean gap ratio $\braket{r}$ close to $0.536$,  exhibiting a gap ratio distribution like Fig. \ref{SFigure_LevelStatistics}(g) which is consistent with the Gaussian orthogonal ensemble (GOE) \cite{atas2013,Giraud2022Probing}. This is consistent with the existence of an antiunitary symmetry of our model, namely complex conjugation in the $S_z$ basis, or the so-called spinless time-reversal symmetry.  This result reveals that most charge sectors are chaotic. There is no additional symmetry that could induce further block diagonalization within these charge sectors. 

However, as shown in Figs. \ref{SFigure_LevelStatistics} (a) and (b), there are a few PBC charge sectors whose mean gap ratio significantly deviates from the GOE value 0.536. Three typical gap ratio distributions shown in Fig. \ref{SFigure_LevelStatistics}(d-f) demonstrate that the energy spectrum of these anomalous PBC charge sectors has many degenerate eigenvalues, which implies the existence of additional symmetry besides the exponential symmetry. For comparison, such anomalous charge sectors are not observed under open boundary conditions [see Fig. \ref{SFigure_LevelStatistics}(c)], where the deviation of $\braket{r}$ near $Q/(2^L-1)\approx \pm 5$ merely comes from the smallness of the Hilbert space dimension for these OBC charge sectors.

The difference between the boundary conditions indicates that the translation symmetry under periodic boundary conditions may play a role in shaping the level statistics in these anomalous PBC charge sectors. In the rest of this section, we focus on the PBC and investigate the role of translation symmetry. As discussed in the main text, the PBC exponential charge operator $\hat Q=\sum_{j=1}^L 2^{L-j} \hat{S}_j^z  \mod (2^L-1)$ and the PBC translation symmetry $\hat T$ have a commutation relation $\hat T \hat Q \hat T^{-1}=2\hat Q \mod (2^L-1)$. Generally, $\hat T$ transfers a state $\ket{\psi_{2Q}}$ in the $2Q$ charge sector satisfying $\hat Q\ket{\psi_{2Q}}=2Q\ket{\psi_{2Q}}$ into another state $\hat T\ket{\psi_{2Q}}$ in the $Q$ charge sector. Here, the eigenvalue of $\hat Q$ is defined mod $2^L - 1$.  This can be easily proved by the observation that $\hat Q\hat T\ket{\psi_{2Q}}=\frac{1}{2}\hat T\hat Q\ket{\psi_{2Q}}=Q\hat T\ket{\psi_{2Q}}$. Therefore, a generic PBC charge sector with the exponential charge $Q$ is likely not invariant under the one-time action of the translation symmetry. 

Nevertheless, the above analysis does not exclude the possibility that for some specific charge sector $Q$ and for an integer $m\in\{1,2,\cdots,L-1\}$  \footnote{$m=0$ and $m=L$ are trivial since $\hat T^0=\hat T^L=I$ is the identity.}, the relation $Q=2^mQ\mod 2^L-1$ can be satisfied. In such a case, the $Q$ sector is invariant under the action of  $\hat T^m$.  As a result,  the $m$-step translation operation $\hat T^m$ can further block diagonalize the $Q$ sector into different Krylov subspaces labeled by different momenta. We remark that these Krylov subspaces are reminiscent of quantum fragmentation, since a general root state with a definite momentum does not have a product-state structure in the computational basis. 

A trivial example is $Q=0$ and $m=1$ for an arbitrary system size $L$, as observed in Fig. \ref{SFigure_LevelStatistics}(a) and (b). We note that Fig. \ref{SFigure_LevelStatistics}(a)  also indicates several anomalous charge sectors with $Q\ne0$ for $L=6$. These charge sectors can be divided into two groups: (1) $Q=\pm 9,\pm18,\pm27$ and $m=3$, as exemplified in Fig. \ref{SFigure_LevelStatistics}(e); (2) $Q=\pm 21$ and $m=2$, as exemplified in Fig. \ref{SFigure_LevelStatistics}(f).

We have revealed that these anomalous charged sectors have an additional symmetry as some subgroups of the translation symmetry. With this intuition in mind, we now discuss the general approach for identifying these anomalous charged sectors in a generic system. With the system size $L$ and an integer $m\in\{1,2,3,\cdots,L-1\}$, we first define the greatest common divisor (gcd) between $m$ and $L$,  i.e., $g=\gcd(m,L)$. Then the equation $Q=2^mQ\mod 2^L-1$ is solved by $Q=0\mod \frac{2^L-1}{2^g-1}$ \footnote{This solution can be proved as follows. First, the original equation $Q=2^mQ\mod 2^L-1$ can also be expressed as $(2^m-1)Q=0\mod 2^L-1$. We can define an integer $g_0=\gcd(2^m-1,2^L-1)$, such that $(2^m-1)Q=0\mod 2^L-1$ is transformed into $\frac{2^m-1}{g_0}Q=0\mod \frac{2^L-1}{g_0}$. Since $\gcd(\frac{2^m-1}{g_0},\frac{2^L-1}{g_0})=1$, we conclude that $Q=0\mod \frac{2^L-1}{g_0}$. Now we need to prove that $g_0=\gcd(2^m-1,2^L-1)=2^{\gcd{(m,L)}}-1=2^g-1$. On the one hand, $g=\gcd(m,L)$ indicates that  $m/g$ and $L/g$ are integers. In this sense, $2^m-1=(2^g)^{m/g}-1=(2^g-1)[1+2^g+(2^g)^2+\cdots+(2^g)^{m/g-1}]$ and  $2^L-1=(2^g)^{L/g}-1=(2^g-1)[1+2^g+(2^g)^2+\cdots+(2^g)^{L/g-1}]$. This tells that $2^g-1$ is a common divisor of $2^m-1$ and $2^L-1$. On the other hand, $\operatorname{gcd}\left(2^m-1,2^L-1\right)=\operatorname{gcd}\left(2^m-1,2^L-1-2^{L-m}\left(2^m-1\right)\right)=\operatorname{gcd}\left(2^m-1,2^{L-m}-1\right)$. This relation further leads to $\operatorname{gcd}\left(2^m-1,2^L-1\right)=\operatorname{gcd}\left(2^m-1,2^{L \mod m}-1\right)$. The change of exponents in these equations is equivalent to the Euclidean algorithm in calculating $\gcd(m,L)$.  As a result, we prove that $\gcd(2^m-1,2^L-1)=\gcd(2^{\gcd{(m,L)}}-1,2^0-1)=2^{\gcd{(m,L)}}-1$. }. There are $2^g-1$ solutions: $Q=\frac{2^L-1}{2^g-1}k$ with $k=0,1,\cdots,2^g-2$. Specifically, when $g=\gcd(m,L)=1$, the only solution is $Q=0$.  As a benchmark, we apply this general result to Fig. \ref{SFigure_LevelStatistics}(a) and (b). In Fig. \ref{SFigure_LevelStatistics}(a), we have $L=6$. First, $\gcd(1,6)=\gcd(5,6)=0$ leads to the $Q=0$ sector; second,  $\gcd(2,6)=\gcd(4,6)=2$ results in three sectors $Q=0,21,42$; last, $\gcd(3,6)=3$ provides seven sectors $Q=0,9,18,27,36,45,54$. In Fig. \ref{SFigure_LevelStatistics}(b), we have $L=5$ and therefore $\gcd(1,5)=\gcd(2,5)=\gcd(3,5)=\gcd(4,5)=1$. The only solution in this case is $Q=0$, as observed in Fig. \ref{SFigure_LevelStatistics}(b).

The above solutions of the anomalous charge sectors have an intuitive picture. As the exponential charge $Q$ is defined up to multiples of $2^L-1$, each $Q$ can be uniquely mapped to a length-$L$ bit string, which can be viewed as a root state. For example, when $L=6$, $Q=0,9,21$ can be mapped into $000000$, $001001$, and $010101$, respectively. Therefore, $g=\gcd(m,L)$, as a factor of $L$, indicates the sub-$L$ periodicity of these bit strings with a periodic boundary condition. This also indicates that there exists only a small number of anomalous charge sectors when the system size $L$ is a composite number. These anomalous sectors have additional symmetries originating from the subgroups of translation. In contrast,  the bit strings for most PBC charge sectors do not have such a periodicity, which indicates a chaotic behavior in their level statistics.

We further note that the level-spacing statistics in the $Q$ and $-Q$ sectors are identical. This result can be understood in terms of a combined symmetry transformation $\hat W=\hat U_z\hat U_x$ with $\hat U_z=\prod_j e^{-i\pi \hat S_j^z}$ and $\hat U_x=\prod_j e^{-i\pi \hat S_j^x}$. We write the Hamiltonian in Eq. \eqref{seq:spin_ham} as $\hat H(J,h,\lambda)$. It is straightforward to check that the unitary transformation $\hat W$ leads to  $\hat W \hat H(J,h,\lambda) \hat W^{-1}=-\hat H(J,h,\lambda)$, while simultaneously mapping the charge operator $\hat Q$ into $\hat W\hat Q \hat W^{-1}=-\hat Q$. As a result, the Hamiltonian restricted to the $-Q$ sector is unitarily related to minus the Hamiltonian in the $Q$ sector. Their spectra are therefore related by $E\to -E$, which explains their identical level statistics.

\begin{figure}[t]
    \centering
    \includegraphics[width=\linewidth]{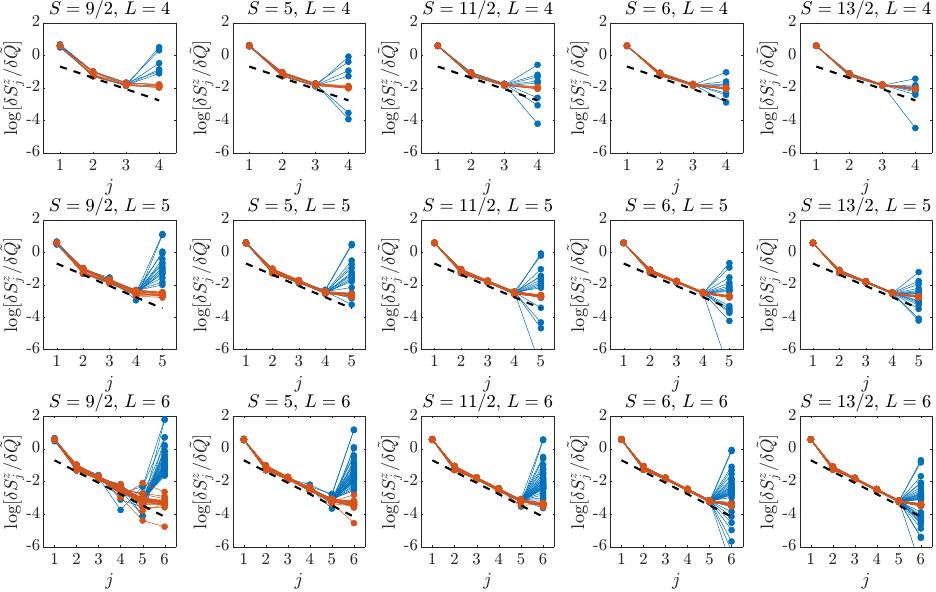}
\caption{Open-boundary distribution of the rescaled local magnetization difference $\delta S_j^z/\delta \tilde Q$, defined in Eq.~\eqref{seq:rescaled_magnetization}. Each panel corresponds to a given pair of $S$ and $L$, and each curve represents one of the symmetry-resolved ground states in several charge sectors $Q$. Denoting by $E_{\mathrm{gs}}(Q)$ the ground-state energy in sector $Q$, and by $E_{\mathrm{gs}}(Q_0)=\min_Q E_{\mathrm{gs}}(Q)$ the global ground-state energy, we show the results for the $2^L-2$ charge sectors with the lowest $E_{\mathrm{gs}}(Q)$ above $E_{\mathrm{gs}}(Q_0)$. The orange (blue) curves correspond to even (odd) charge difference $Q-Q_0$. The dashed line indicates the scaling $\delta S_j^z/\delta \tilde Q\sim 2^{-j}$. Parameters are set to $J=2$ and $h=\lambda=1$. }
\label{SFigure_DeltaSz}
\end{figure}
\begin{figure}[tb]
    \centering
    \includegraphics[width=\linewidth]{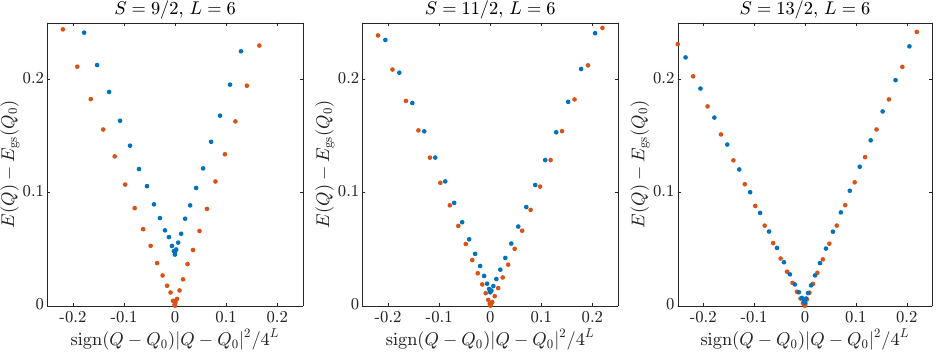}
    \caption{Low-energy OBC spectrum for the spin value $S$ indicated above each plane, where $E_\text{gs}(Q)$ denotes the ground-state energy in charge sector $Q$, and $Q_0$ labels the charge sector of the global ground state. The orange (blue) points correspond to even (odd) charge difference $Q-Q_0$. Parameters are set to $L=6$, $J=2$, and $h=\lambda=1$.
    }
    \label{SFigure_energyband}
\end{figure}
\section{Magnetization distribution under open boundary conditions}

In this section, we present additional numerical results under open boundary conditions for the local magnetization difference between the low-energy states and the corresponding global ground state [see the inset of Fig. 2(f) in the main text], which reflects the presence of the left edge mode. This is quantified by the density deviation $\delta S_j^z$ on the site $j$:
\begin{equation}\label{seq:magnetization}
    \delta S_j^z=|\braket{\psi_Q|\hat{S}_j^z|\psi_Q}-\braket{\psi_{Q_0}|\hat{S}_j^z|\psi_{Q_0}}|,
\end{equation}
where $\ket{\psi_Q}$ denotes the ground-state wave function in the OBC charge sector $Q$, and $Q_0$ labels the charge sector with the lowest ground-state energy under open boundary conditions. Namely, $E_{\mathrm{gs}}(Q_0)=\min_Q E_{\mathrm{gs}}(Q)$ with $E_{\mathrm{gs}}(Q)$ being the ground-state energy in the charge sector $Q$. To better illustrate our results, we define a rescaled local magnetization difference $\delta\tilde{S}_j^z$ as
\begin{equation}\label{seq:rescaled_magnetization}
    \delta\tilde{S}_j^z\equiv\frac{\delta S_j^z}{ \delta \tilde Q}=\frac{2^L\delta S_j^z}{|Q-Q_0|}
\end{equation}
where $
\delta\tilde{Q}=\frac{|Q-Q_0|}{2^L}$ is the rescaled global charge difference.

The numerical results for $\delta\tilde S_j^z\equiv \delta S_j^z/\delta\tilde Q$ at various $S$ and $L$ are shown in Fig.~\ref{SFigure_DeltaSz}. The orange and blue curves correspond to even and odd charge differences $Q-Q_0$, respectively. We find that, for different low-energy modes, the bulk profiles of $\delta\tilde S_j^z$ decay exponentially away from the left edge at $j=1$. These profiles collapse onto the common scaling $\delta\tilde S_j^z\sim 2^{-j}$, with the collapse becoming more accurate for larger $S$ and $L$. The observed scaling collapse indicates that $\delta S_j^z$ is approximately described by  $\delta S_j^z \sim 2^{-j}(Q-Q_0)/2^L$. As discussed in the main text, this observation is compatible with the effective edge Hamiltonian $\hat{\mathcal H}_{\mathrm{edge}}\propto \frac{g_0}{2}(\partial_t\delta\hat\theta_{\mathrm{edge}})^2
\propto [(\hat Q-Q_0)/2^L]^2$, obtained by quantizing the zero mode $\delta\theta_j=2^{-j}\delta\theta_{\mathrm{edge}}$ in the low-energy effective Lagrangian of phase fluctuations. The resulting quadratic low-energy dispersion in $(Q-Q_0)/2^L$ is also confirmed numerically, with examples shown in Figs.~\ref{SFigure_energyband} and \ref{SFigure_OBC_Thermal}(f).

The exponential decay is violated only near the rightmost site $j=L$, where the profiles exhibit a pronounced dependence on the parity of the charge difference $Q-Q_0$ (orange for even, blue for odd). Within the same energy window, comparing the highest blue point at $j=L$ for different $S$ at fixed $L$, we find that this right-edge deviation becomes progressively weaker with increasing $S$, indicating that it originates from finite-$S$ quantum fluctuations and becomes negligible in the large-$S$ limit.

We note that the charge parity-dependent profile near the right boundary stems from a right-boundary $\mathbb{Z}_2$ symmetry in the OBC quantum breakdown model. Since $\hat Q=\sum_{j=1}^L 2^{L-j}\hat {S}_j^z$, the charge parity operator can be rewritten as
\begin{equation}\hat P=e^{i\pi \hat Q}=e^{i\pi \hat S_L^z}.\end{equation}
Namely, it is solely determined by the parity of spin $\hat S_L^z$ of the last site $L$. Since $\hat Q$ is the symmetry charge, one can straightforwardly verify that $[\hat P,\hat H]=0$ for the OBC Hamiltonian $\hat H$ in Eq.~\eqref{seq:spin_ham}. Intuitively, this symmetry $\hat P$ is because $\hat H$ can only change $\hat S_L^z$ by multiples of $2$. Therefore, if $Q-Q_0$ is even (odd), the charge sector $Q$ can access the same (opposite) $\hat S_L^z$-parity sub-Hilbert space on site $L$ as the charge sector $Q_0$, exhibiting an even-odd effect. This leads to charge parity dependent values of $\braket{\psi_Q|\hat S_L^z|\psi_Q}$ at the right boundary site [Fig. \ref{SFigure_DeltaSz}, orange for even, blue for odd]. The same parity dependence also appears in the low-energy OBC spectrum in Fig.~\ref{SFigure_energyband}, where the two branches of symmetry-resolved ground-state energies correspond to different parities of $Q-Q_0$. The $\braket{\psi_Q|\hat S_L^z|\psi_Q}$ fluctuation on the right-most site $L$ is thus not due to a boson edge mode. We further note that this energy splitting decreases with increasing $S$, similar to the decrease of the PBC ground-state bandwidth $W$ with $S$ shown in Fig.~2(h) of the main text.

\section{Ground state phase transition and finite-temperature phase diagram}\label{sec:thermal}
\begin{figure}[t]
    \centering
    \includegraphics[width=\linewidth]{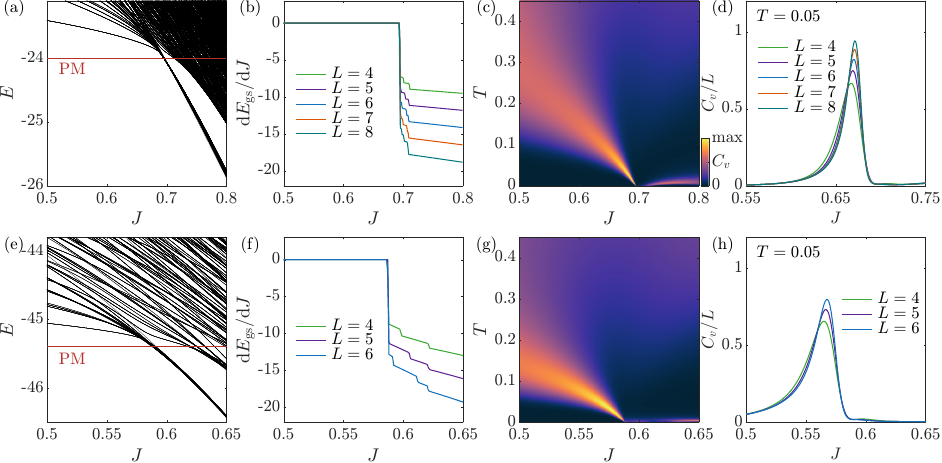}
    \caption{(a,e) The PBC low-energy ED spectrum $E$ with respect to the breakdown interaction $J$. The red line with label PM represents the energy of the paramagnetic (PM) state. A clear level crossing between PM and $2^L-1$ SSB ground states indicates a first order phase transition. (b,f) The ground-state energy derivative $\mathrm{d}E_\mathrm{gs}/\mathrm{d}J$. (c,g) The finite-temperature colormap representing the heat capacity $C_v$, which agrees with the expected phase diagram in the main text. (d,h) The specific heat $C_v/L$ with $T=0.05$ for different system sizes. In  (a-d), we fix $S=3,\ \lambda=h=1$. We also fix $L=8$ for (a) and (c). In  (e-h), we fix $S=5,\ \lambda=0.5,\ h=1$. We also fix $L=6$ for (e) and (g). We take PBC in all plots here.}
    \label{SFigure_Thermal}
\end{figure}

\begin{figure}[t]
    \centering
    \includegraphics[width=\linewidth]{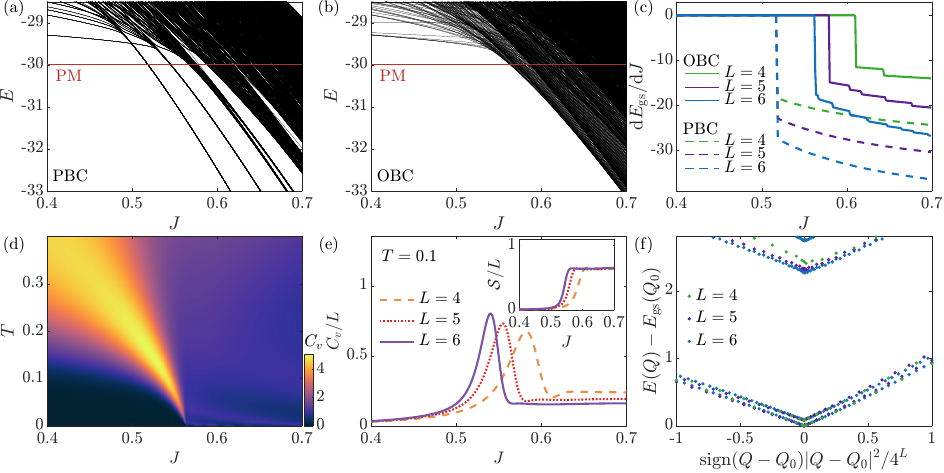}
    \caption{The ED spectra and thermal quantities under different boundary conditions for $S=5, \lambda=h=1$ (the parameters of the main example in the main text). (a) The low-energy PBC spectrum $E$ with respect to the breakdown interaction $J$ for the system size $L=6$. (b) The low-energy OBC spectrum $E$ with respect to the breakdown interaction $J$ for the system size $L=6$. The red lines in (a) and (b), which are labeled by PM, represent the energy of the paramagnetic state. (c) The ground-state energy derivative $\mathrm{d}E/\mathrm{d}J$ for PBC (dashed lines labeled by the system size $L$) and OBC (solid lines labeled by the system size $L$) systems. (d) The heat capacity $C_v$ for $L=6$ under open boundary conditions. (e) The specific heat $C_v/L$ and the entropy density $\mathcal{S}/L$ (inset) at $T=0.1$ for different system sizes $L$ under open boundary conditions. (f) The low-energy OBC spectrum for $L=6$ and $J=1$. }
    \label{SFigure_OBC_Thermal}
\end{figure}

In this section, we provide additional data for the zero-temperature phase transition and finite-temperature crossover, for which the ED numerical results as presented in Fig. \ref{SFigure_Thermal} and Fig. \ref{SFigure_OBC_Thermal}.

The zero-temperature first-order phase transition between the two distinct types of ground states is revealed by a level crossing between the paramagnetic state and the SSB manifold, as shown in Figs. \ref{SFigure_Thermal}(a), \ref{SFigure_Thermal}(e), and \ref{SFigure_OBC_Thermal}(a) for PBC. This level crossing is further manifested by discontinuous jumps in the derivative of the ground-state energy with respect to the breakdown interaction strength, as displayed in Figs. \ref{SFigure_Thermal}(b), \ref{SFigure_Thermal}(f), and \ref{SFigure_OBC_Thermal}(c). A similar first-order phase transition is also observed in the low-energy spectrum of systems under open boundary conditions (OBC), as shown in Figs. \ref{SFigure_OBC_Thermal}(b) and \ref{SFigure_OBC_Thermal}(c).  We remark that the PBC phase transition points [Figs. \ref{SFigure_Thermal}(b) and \ref{SFigure_Thermal}(f)] are independent of the system size $L$. In contrast, the OBC phase transition points [Fig. \ref{SFigure_OBC_Thermal}(c)] exhibit a finite-size effect, converging to the PBC transition points as the system size $L$ increases. This is also confirmed by the large-scale DMRG simulations presented in the End Matter.

The zero-temperature phase transition from PM to the SSB quantum breakdown condensate is expected to remain a sharp crossover in the low-temperature regime. Due to the sharp difference in ground-state degeneracy between two phases, the zero-temperature phase transition and finite-temperature crossover can be estimated by entropy considerations.

Subsequently, we present more details of this low temperature entropy calculation. For $J$ near the SSB phase transition point $J_0$, the low energy competing states are the symmetric PM ground state with energy $E_\mathrm{gs}^\mathrm{sym}$, and the SSB ground state in each charge sector $Q$ with energy $E_\mathrm{gs}(Q)$. Assume the state $E_\mathrm{gs}(Q_0)$ in charge sector $Q_0$ is the lowest among all the SSB ground states. The ED results indicate that [Figs. \ref{SFigure_Thermal}(a) and \ref{SFigure_Thermal}(e)] \begin{equation}
E_\mathrm{gs}^\mathrm{sym}-E_\mathrm{gs}(Q_0)=v_\mathrm{gs}(J-J_0)L\ ,\quad (v_\mathrm{gs}>0)
\end{equation}
which effectively describes the level crossing at $J=J_0$. At low temperatures $T=1/\beta$ (we set $k_B=1$), to the leading order we ignore all the states above the bulk gap $\Delta$, and obtain an approximate partition function given by keeping only the above competing ground states:
\begin{equation}
Z=e^{-\beta E_\mathrm{gs}^\mathrm{sym}}+\sum_Q e^{-\beta E_\mathrm{gs}(Q)}\ .
\end{equation}

(i) For PBC, there are $2^L-1$ charge $Q$ sectors, of which the ground state energies $E_\mathrm{gs}(Q)$ lie in the energy interval $[E_\mathrm{gs}(Q_0),E_\mathrm{gs}(Q_0)+W]$, where $W$ is of order one ($W<\Delta$). For convenience, we approximately regard $E_\mathrm{gs}(Q)$ as uniformly distributed in the bandwidth $W$ (we note that the conclusion does not change even if $E_\mathrm{gs}(Q)$ is not uniformly distributed, as long as $W$ is finite), we find 
\begin{equation}
\begin{split}
Z&\approx e^{-\beta E_\mathrm{gs}^\mathrm{sym}}+ (2^L-1)e^{-\beta E_\mathrm{gs}(Q_0)}\int_0^W dE \frac{e^{-\beta E}}{W}\\
&\approx e^{-\beta E_\mathrm{gs}(Q_0)}\left[e^{-\beta v_\mathrm{gs}(J-J_0)L}+2^L\frac{1-e^{-\beta W}}{\beta W}\right]\ .
\end{split}
\end{equation}

(ii) For OBC, in ED (and by the analytical edge mode argument in the main text) we approximately find $E_\mathrm{gs}(Q)-E_\mathrm{gs}(Q_0)\approx D\left(\frac{Q-Q_0}{2^L}\right)^2$. An example can be found in Fig. \ref{SFigure_OBC_Thermal}(f) [see also Fig. 2(d) of the main text]. Thus,
\begin{equation}
\begin{split}
Z&\approx e^{-\beta E_\mathrm{gs}^\mathrm{sym}}+ e^{-\beta E_\mathrm{gs}(Q_0)}\sum_Q e^{-D\left(\frac{Q-Q_0}{2^L}\right)^2}\\
&\approx e^{-\beta E_\mathrm{gs}(Q_0)}\left[e^{-\beta v_\mathrm{gs}(J-J_0)L}+2^L\int_{-\infty}^{\infty} dx e^{-\beta D x^2}\right]\\
&= e^{-\beta E_\mathrm{gs}(Q_0)}\left[e^{-\beta v_\mathrm{gs}(J-J_0)L}+2^L\sqrt{\frac{\pi}{\beta D}}\right]\ ,\\
\end{split}
\end{equation}
where we have defined $x=\frac{Q-Q_0}{2^L}$ in the second line.

It is then easy to calculate the thermal entropy $\mathcal{S}=\ln Z-\beta\partial_\beta\ln Z$. Interestingly, in the $L\rightarrow\infty$ limit, the entropy density $\mathcal{S}/L$ for both PBC and OBC takes the same form:
\begin{equation}
\frac{\mathcal{S}}{L}=
\begin{cases}
&0\ ,\qquad\quad (J<J_T) \\
&\ln 2\ ,\qquad (J\ge J_T)
\end{cases}\ ,\qquad J_T=J_0-v_\mathrm{gs}^{-1}T\ln2\ .
\end{equation}
This result further supports that the difference between PBC and OBC is only on the boundaries, not in the bulk, as is expected for physical systems with local Hamiltonians. This implies that the crossover boundary $J_T$ at finite temperature is on the $J<J_0$ side, in agreement with our ED calculations.

According to the Landau-Peierls instability, and the more generic theorem in Ref. \cite{ruelle1969statistical} for 1D local Hamiltonians with finite on-site Hilbert spaces, at finite temperature $T>0$, thermal fluctuations from proliferating domain walls, which are excitations above the bulk gap, will always restore the SSB and forbids any phase transitions. Therefore, we expect $J=J_T$ to be a crossover boundary at $T>0$.

The finite-temperature colormaps of heat capacity shown in Figs.\ref{SFigure_Thermal}(c),  \ref{SFigure_Thermal}(g), and \ref{SFigure_OBC_Thermal}(d) resemble that in Fig. 3(b) of the main text, indicating a finite-temperature crossover realizing a Pomeranchuk effect. Near the crossover boundary, the specific heat $C_v/L$ shows a sharp peak [Figs.\ref{SFigure_Thermal}(d), \ref{SFigure_Thermal}(h), and \ref{SFigure_OBC_Thermal}(e)], which tends to sharpen for the larger system size $L$ calculated, indicating a sharp crossover at low temperatures. The finite-size effect of the zero-temperature phase transition for OBC systems is also observed for the finite-temperature crossover. As shown in Fig. \ref{SFigure_OBC_Thermal}(e), the locations of specific heat peaks and entropy jumps move monotonically with respect to the system size $L$, showing a considerable finite size effect.

We note that while Fig.~\ref{SFigure_OBC_Thermal} is calculated with parameters the same as that in Fig. 3 of the main text, Fig.~\ref{SFigure_Thermal} presents two different sets of parameters. These results demonstrate that the zero-temperature first-order transition and finite-temperature crossover remain robust against variations of the microscopic model details.

\section{Additional data on glassy dynamics}
\begin{figure}
    \centering
    \includegraphics[width=\linewidth]{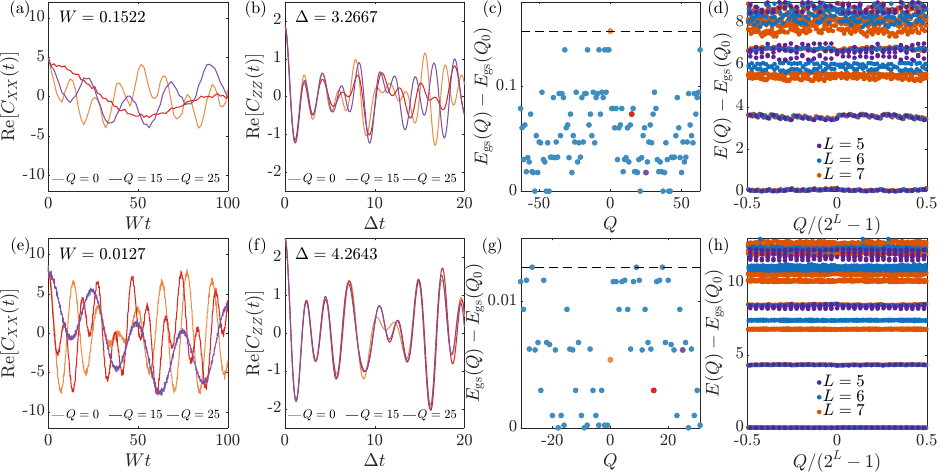}
    \caption{(a,e) The autocorrelation functions $C_{XX}(t)$ calculated by ED. (b,f) The autocorrelation functions $C_{ZZ}(t)$ calculated by ED. (c,g) The energy distribution of the PBC ground-state manifold, where the dashed line labels the bandwidth $W$. (d,h) The low-energy PBC spectra for different system sizes. We set $S=3,\ \lambda=0.5,\  h=1,\ J=2$ in (a-d) and $ L=7$ in (a-c). Similarly, the parameters in (e-h) are $S=4,\ \lambda=1.0,\  h=1,\ J=2$, with $L=6$ in (e-g).} 
    \label{SFigure_Autocorrelation}
\end{figure}

\begin{figure}
    \centering
    \includegraphics[width=\linewidth]{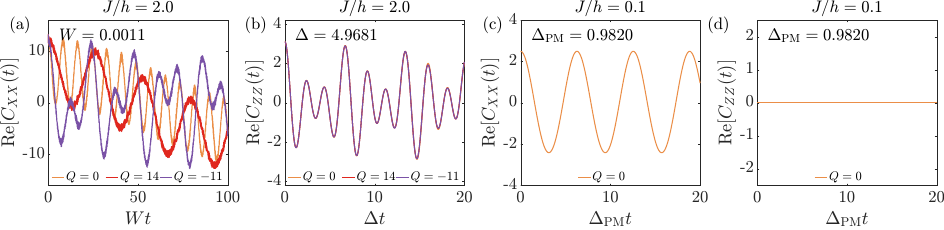}
    \caption{The autocorrelation functions $C_{XX}(t)$ and $ C_{ZZ}(t)$ in the PBC system with $S=5$ and $L=5$. (a,b) show the results in the SSB phase ($J=2$ and $h=\lambda=1$) for several charge sectors.  (c,d) present the results in the symmetric phase ($J=0.1$ and $h=\lambda=1$) with the global ground state living in $Q=0$ sector. $W$ in (a) represents the energy bandwidth of the SSB ground-state manifold, and $\Delta$ in (b) denotes the excitation gap between this nearly degenerate ground-state manifold and other high-energy excited states. $\Delta_{\text{PM}}$ in (c,d) shows the energy gap above the unique ground state in the paramagnetic phase. }
    \label{SFigure_TwoPhaseAutocorrelation}
\end{figure}
In this section, we further investigate the glassy dynamics via ED [Figs. \ref{SFigure_Autocorrelation} and \ref{SFigure_TwoPhaseAutocorrelation}].  For a given spin chain with PBC, we study the autocorrelation functions with the initial state $\ket{\psi_Q}$ being the ground state of a specific charge sector $Q$. Then we evaluate the autocorrelation function  
\begin{equation}
    C_{XX}(t)=\braket{\psi_Q|\hat{S}_{L}^x(t)\hat{S}_{L}^x(0)|\psi_Q}, \quad  C_{ZZ}(t)=\braket{\psi_Q|\hat{S}_{L}^z(t)\hat{S}_{L}^z(0)|\psi_Q}-\braket{\psi_Q|\hat{S}_{L}^z(t)|\psi_Q}\braket{\psi_Q|\hat{S}_{L}^z(0)|\psi_Q}.
\end{equation}
Here, the time evolution of an operator $\hat{O}$ is given by $\hat{O}(t)=e^{i\hat{H}t}\hat{Q}e^{-i\hat{H}t}$. Without loss of generality for PBC, we focus on the spin operators on the last site $j=L$. The numerical results for the autocorrelation functions on other sites exhibit similar behaviors.   

As discussed in the End Matter, in the SSB phase under periodic boundary conditions, the timescale of $C_{XX}(t)$ is set by $\tau_X\propto 1/W$, where $W$ is the energy bandwidth of the PBC ground-state manifold [Figs.~\ref{SFigure_Autocorrelation}(c,g)]. By contrast, the timescale of $C_{ZZ}(t)$ is set by $\tau_Z\propto 1/\Delta$, where $\Delta$ is the energy gap above the nearly degenerate ground-state manifold [Figs.~\ref{SFigure_Autocorrelation}(d,h)]. In large spin systems, we have $\Delta\gg W$ [see Fig. 2(h) of the main text], which separates the timescale of the two autocorrelation functions. This analytical expectation is further supported by ED results for other spin values $S$ and model parameters: Figs.~\ref{SFigure_Autocorrelation}(a,e) and \ref{SFigure_TwoPhaseAutocorrelation}(a) show the slow dynamics of $C_{XX}(t)$, while Figs.~\ref{SFigure_Autocorrelation}(b,f) and \ref{SFigure_TwoPhaseAutocorrelation}(b) show the fast dynamics of $C_{ZZ}(t)$.

Furthermore, Figures~\ref{SFigure_TwoPhaseAutocorrelation}(c) and \ref{SFigure_TwoPhaseAutocorrelation}(d) show the ground-state autocorrelation functions in the paramagnetic phase. In this regime, the system has a unique symmetric ground state---the paramagnetic state---in the charge sector $Q=0$. As shown in Fig.~\ref{SFigure_TwoPhaseAutocorrelation}(c), $C_{XX}(t)$ exhibits rapid oscillations on a timescale set by the $O(1)$ gap $\Delta_{\mathrm{PM}}$ between the paramagnetic ground state and the excited states. By contrast, Fig.~\ref{SFigure_TwoPhaseAutocorrelation}(d) shows that $C_{ZZ}(t)=0$, which comes from the fact that the paramagnetic ground state used as the initial state is an eigenstate of both the quantum breakdown Hamiltonian and the local operator $S_j^z$. These results further support that the glassy dynamics is closely tied to the spontaneous breaking of the exponential symmetry under periodic boundary conditions.

\section{Perturbation of symmetry-breaking fields}
\begin{figure}
    \centering
    \includegraphics[width=\linewidth]{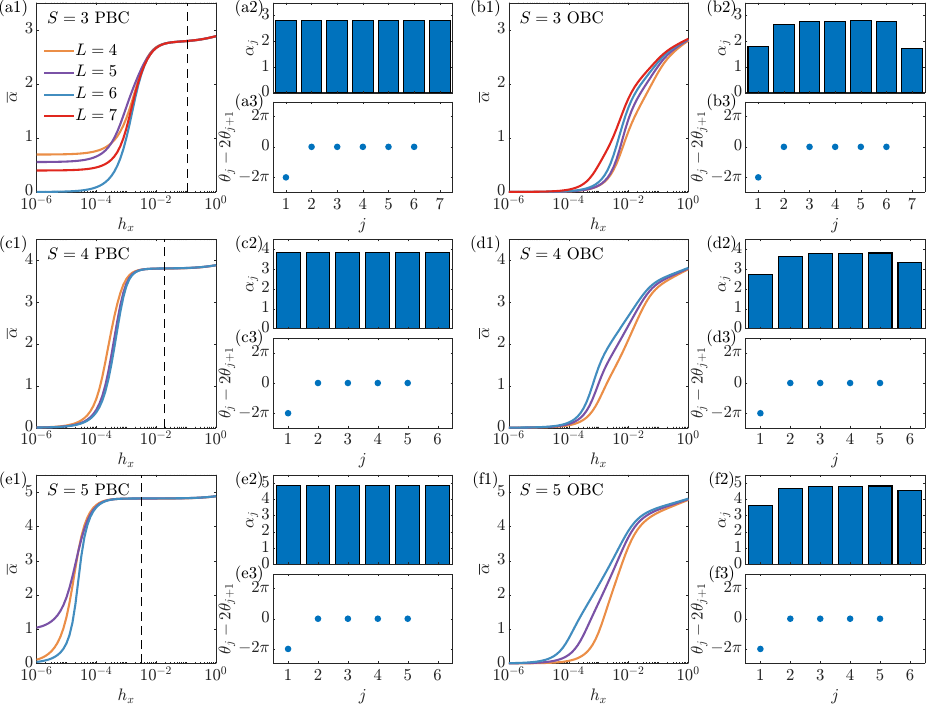}
\caption{Ground-state properties of the perturbed breakdown model in Eq. \eqref{seq:pert_Ham}. The first and second columns show ED results under periodic boundary conditions, while the third and fourth columns show the corresponding results under open boundary conditions. Each row corresponds to a specific choice of $S\in\{3,4,5\}$. The first and third columns present the order parameter $\overline{\alpha}$ [Eq. \eqref{seq:alpha}] for different system sizes $L$ and perturbation strengths $h_x$. The dashed line in the first column indicates the energy bandwidth $W$ of the PBC ground-state manifold at the largest accessible system size. The second and fourth columns show a representative ground-state in-plane spin configuration $\alpha_j$ in Eq. \eqref{seq:alpha_j}, 
where we fix $h_x=0.1$. Other parameters are set to $J=h=\lambda=1$.
}    \label{SFigure_perturbation}
\end{figure}

In this section, we numerically investigate the following symmetry-breaking field perturbed breakdown Hamiltonian:
\begin{equation}\label{seq:pert_Ham}
\begin{split}
   \hat H_{\text{pert}}=&-\frac{J}{(2 S)^{3 / 2}}\sum_{j=1}^{L_{P\backslash O}} \left[\hat {S}_{ j}^{-}(\hat {S}_{ j+1}^+)^2 +\hat {S}_{ j}^{+}(\hat {S}_{ j+1}^{-})^2 \right]-  h\sum_{j=1}^L\left[\lambda \hat {S}_{j}^z+(1-\lambda)\frac{4 S^2}{\pi^2}\sin \left(\frac{\pi}{2S} \hat {S}_{ j}^z\right)\right]\\\
    &-h_x\sum_{j=1}^L\left[\cos(2^{L-j}\phi_L)\hat S_j^x+\sin(2^{L-j}\phi_L)\hat S_j^y\right].
    \end{split}
\end{equation}
The first line is the same as the quantum breakdown Hamiltonian in Eq. \eqref{seq:spin_ham}. The second line is a symmetry-breaking perturbation $h_x$ with a chosen angle $\phi_L=2\pi/(2^L-1)$. In general, $\phi_L$ can be chosen in any of the $2^L-1$ directions corresponding to the $\mathbb{Z}_{2^L-1}$ symmetry (in the discussion of the End Matter we have chosen $\phi_L=0$ instead). The second line explicitly breaks the exponential symmetry under either periodic or open boundary conditions, selecting a specific symmetry-breaking direction. In this section, we employ exact diagonalization to find the global ground state $\ket{\Psi(h_x)}$ of $\hat H_{\text{pert}}$ and calculate the expectation values of $\hat S_j^+ = \hat S_j^x+i\hat S_j^y$:
\begin{equation}\label{seq:alpha_j}
    \braket{\Psi(h_x)|\hat S_j^+|\Psi(h_x)}\equiv \alpha_je^{i\theta_j},
\end{equation}
where the complex value has the amplitude $\alpha_j$ and phase $\theta_j$. 

In all simulated cases with a moderate $h_x$ under either periodic or open boundary conditions, we find that the global ground state has $\theta_j=2^{L-j}\phi_L \mod 2\pi$ independent of $h_x$. Accordingly, the relation $\theta_j-2\theta_{j+1}=0\mod 2\pi$ is satisfied [see the second and fourth  columns of Fig. \ref{SFigure_perturbation} for representative examples]. These results indicate that the spatially dependent symmetry-breaking field $h_x$ in $H_{\text{pert}}$ picks a specific direction labeled by $\phi_L=\frac{2\pi}{2^L-1}$ in the (nearly) degenerate SSB ground-state manifold under periodic boundary conditions with $\phi_L=\frac{2\pi}{2^L-1}\mathbb{Z}_{2^L-1}$ or open boundary conditions with $\phi_L\in[0,2\pi)$. 

To further characterize the effect of the perturbation, we define the averaged in-plane order parameter
\begin{equation}\label{seq:alpha}
    \overline{\alpha}\equiv \frac{1}{L}\sum_{j=1}^{L}\alpha_j,
\end{equation}
the numerical results of which for different spin values $S$ and system sizes $L$ are shown in the first and third columns of Fig.~\ref{SFigure_perturbation}. Like in the End Matter, the ED results here shown in Fig. \ref{SFigure_perturbation} also leads to the same conclusion: under both periodic and open boundary conditions, the dependence of $\overline{\alpha}$ on the symmetry-breaking field strength $h_x$ shows a tendency that
\begin{equation}
    \lim_{h_x\to 0^+}\lim_{L\to\infty}\overline{\alpha}\neq 0.
\end{equation}
Therefore, in the thermodynamic limit, the spontaneous breaking of the exponential symmetry can be detected through the response to an infinitesimal symmetry-breaking field. We further note that under periodic boundary conditions, the ground states almost always have exact degeneracy protected by the translation symmetry, except for the $Q=0$ charge sector ground state. This means if the global ground state is in an exactly degenerate ground state subspace, an infinitesimal symmetry breaking term will induce tunnelings among these degenerate states and explicitly breaks the exponential symmetry in the new ground state (forming a dead-cat state), even if the system size $L$ is finite [see examples in the first column of Fig. \ref{SFigure_perturbation}].

\section{Disorder string operators (Patch symmetry operators)}
\begin{figure}[t]
    \centering
    \includegraphics[width=\linewidth]{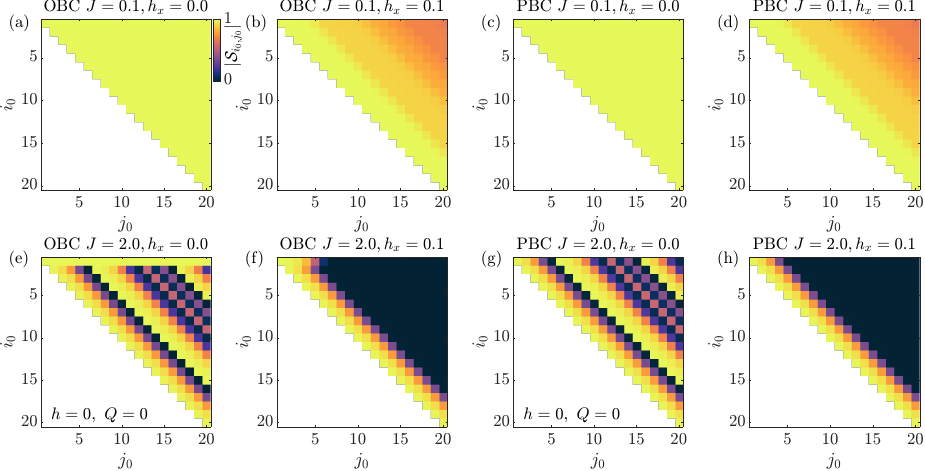}
    \caption{The ground-state expectation value $\mathcal{S}_{i_0,j_0}(\eta)$ for the disorder string operator in Eq. \eqref{seq:string_op} with $\eta=0.1$. We consider both open (a,b,e,f) and periodic (c,d,g,h) boundary conditions. The top row is in the paramagnetic phase ($J=0.1$), and the bottom row is in the SSB phase ($J=2.0$). We set $h=0$ for (e,g) where we employ charge-conserving DMRG to find the ground state in the $Q=0$ charge sector. Other plots have $h=1$. In all plots, the other parameters are set to $S=5$, $\lambda=1$, and $L=20$.} 
    \label{SFigure_String}
\end{figure}

In this section, to further reveal the unconventional nature of the SSB phase of our model, we present additional numerical results for the expectation values of the disorder string operator:
\begin{equation}\label{seq:string_op}
    \mathcal{\hat S}_{i_0,j_0}(\eta)=\exp(i\eta\sum_{n=i_0}^{j_0}2^{j_0-n}\hat S_n^z)
\end{equation}
where $1\le i_0\le j_0\le L$ are two end points of the string operator and $\eta>0$ is a phase parameter. Here $
\hat Q_{i_0,j_0}=\sum_{n=i_0}^{j_0}2^{j_0-n}\hat S_n^z$ is the subsystem generator of the exponential symmetry in a patch of interval $[i_0,j_0]$, which is also known as the \emph{patch symmetry} in the literature of generalized symmetry \cite{Ji2020Categorical,Han2024Topological}.

For conventional SSB phases, such a disorder string operator (patch symmetry operator) creates two domain walls (defects) at sites $i_0$ and $j_0$, and the SSB ground states inside and outside the domain walls will differ. Accordingly, the expectation value of the disorder string operator will decay exponentially with $|j_0-i_0|$ in a conventional SSB phase. In contrast, it will be non-decaying in a conventional symmetric phase.

We employ DMRG to obtain the ground state $\ket{\psi}$ of the Hamiltonian Eq. \eqref{seq:pert_Ham} under both periodic and open boundary conditions, and then calculate the expectation value $\mathcal{S}_{i_0,j_0}(\eta)=\braket{\psi|  \mathcal{\hat S}_{i_0,j_0}(\eta)|\psi}$ for different choices of $i_0$ and $j_0$. The numerical results with $\eta=0.1$ are shown in Fig. \ref{SFigure_String}, where we present the outcomes for both $h_x=0$ and $h_x=0.1$. In the SSB phase with $h_x=0$, we further set $h=0$ [Figs. \ref{SFigure_String}(e) and \ref{SFigure_String}(g)] and use the charge-conserving DMRG to obtain the $Q=0$ sector ground state $\ket{\psi_{Q=0}}$ under both boundary conditions, which belongs to the ground-state manifold when taking PBC and is found to be close to the finite-size global ground state when taking OBC. Other choices of $Q$ and parameters will provide similar results as long as the system is in the SSB phase. In all cases, we find that $|\mathcal{S}_{i_0,j_0}(\eta)|$ in the bulk region only depends on $j_0-i_0$, exhibiting a translation invariance.

Importantly, at $h_x=0$, the disorder string operator remains non-decaying with $|j_0-i_0|$ in both the symmetric phase [Figs.~\ref{SFigure_String}(a) and \ref{SFigure_String}(c)] and the SSB phase [Figs.~\ref{SFigure_String}(e) and \ref{SFigure_String}(g)]. This behavior is in sharp contrast to that of conventional U(1) symmetry, for which the disorder string operator in a finite-size ground state is expected to decay with the separation $|j_0-i_0|$ in the SSB phase. The disorder string operator, therefore, does not provide a reliable diagnostic of spontaneous breaking of the exponential symmetry in finite systems, at least in the simplest setup. This observation further highlights that the SSB phase identified here lies beyond conventional SSB quantum phases.

The disorder string operator becomes decaying only in the presence of an explicit symmetry-breaking perturbation [$h_x\ne0$ in Eq.~\eqref{seq:pert_Ham}], which pins the ground state to a particular SSB pattern. As shown in Figs.~\ref{SFigure_String}(f) and \ref{SFigure_String}(h), $|\mathcal{S}_{i_0,j_0}(\eta)|$ rapidly decreases with the string length $j_0-i_0$. This behavior can also be understood by evaluating the expectation value of the bosonic disorder string operator $\mathcal{\hat S}_{i_0,j_0}^{(\mathrm{b})}(\eta)=\exp\left(i\eta\sum_{n=i_0}^{j_0}2^{j_0-n}\hat a_n^\dagger \hat a_n\right)$ with respect to the exact SSB wave function $|\Psi_{\mathrm{ssb}}(\theta)\rangle=\prod_{j=1}^L\exp\left(\alpha e^{i2^{L-j}\theta}\hat a_j^\dagger\right)|0\rangle$ at the RK point [see Eq.~(5) of the main text], which becomes the global ground state in the presence of a symmetry breaking perturbation $h_x$ with $\phi_L=\theta$. Using the coherent-state identity $\langle z|e^{i\gamma \hat a_n^\dagger \hat a_n}|z\rangle/\langle z|z\rangle=\exp\left[|z|^2(e^{i\gamma}-1)\right]$, we obtain
\begin{equation}
\left|\frac{\braket{\Psi_\text{ssb}(\theta)| \mathcal{\hat S}_{i_0,j_0}^{(\text{b})}(\eta)|\Psi_\text{ssb}(\theta)}}{\braket{\Psi_\text{ssb}(\theta)|\Psi_\text{ssb}(\theta)}}\right|=\exp[-2|\alpha|^2\sum_{n=0}^{j_0-i_0}\sin^2(2^{n-1}\eta)],
\end{equation}
This expression decreases monotonically with $j_0-i_0$ and typically exhibits exponential decay for a generic, non-fine-tuned choice of $\eta$. 

When $h_x$ is nonzero, however, there is no SSB phase transition, and this disorder string operator is decaying near both the original symmetric phase [Fig.~\ref{SFigure_String}(b) and~\ref{SFigure_String} (d)] and the original SSB phase [Figs.~\ref{SFigure_String}(f) and ~\ref{SFigure_String}(h)].

\bibliography{spin_ref}